\documentclass[useAMS,usenatbib]{mn2e}

\newcommand{\lya}{Ly$\alpha$}
\newcommand{\lyman}{Lyman-$\alpha$}
\newcommand{\msun}{\mbox{M}_{\odot}}
\newcommand{\be}{\begin{enumerate}}
\newcommand{\ee}{\end{enumerate}}
\newcommand{\kms}{\,km\,s$^{-1}$}
\newcommand{\rhot}{$T$-$\rho$}
\newcommand{\hmpc}{h^{-1}\,{\rm Mpc}}
\newcommand\ionm[2]{#1$\;${\small\rmfamily{#2}}\relax}% 

\newcommand{\apj}{ApJ}
\newcommand{\apjl}{ApJL}
\newcommand{\apjs}{ApJS}
\newcommand{\aj}{AJ}

\newcommand{\mnras}{MNRAS}

\newcommand{\araa}{ARA\&A}

\usepackage{graphicx}
\usepackage{amssymb}
\usepackage{amsfonts}
\usepackage{verbatim}
\usepackage{times}
\usepackage[usenames]{color}

\begin{document}

\title[Scale of the Longitudinal \lyman\ Forest]{Pressure Support vs.\ Thermal Broadening in the Lyman-$\alpha$ Forest I:\\
Effects of the Equation of State on Longitudinal Structure}
\author[Peeples et al.]{Molly S.\ Peeples$^{1}$\thanks{E-mail:
    molly@astronomy.ohio-state.edu},  David H.\ Weinberg$^{1}$, Romeel
  Dav\'e$^{2}$, Mark A.\ Fardal$^{3}$, Neal Katz$^{3}$\\
$^{1}$Department of Astronomy and the Center for Cosmology and
  Astro-Particle Physics, The Ohio State University, Columbus,~OH~43210\\
$^{2}$University of Arizona, Steward Observatory, Tucson,~AZ~85721\\
$^{3}$Department of Astronomy, University of Massachusetts, Amherst,~MA~01003}

\pagerange{\pageref{firstpage}--\pageref{lastpage}} \pubyear{2009}

\date{\today}

\maketitle

\label{firstpage}

\begin{abstract}
In the low density intergalactic medium (IGM) that gives rise to the
\lyman\ forest, gas temperature and density are tightly correlated.  The
velocity scale of thermal broadening and the Hubble flow across the gas
Jeans scale are of similar magnitude ($H\lambda_J \sim \sigma_{\rm
th}$). To separate the effects of gas pressure support and thermal
broadening on the \lya\ forest, we compare spectra extracted from two
smoothed particle hydrodynamics (SPH) simulations evolved with different
photoionization heating rates (and thus different Jeans scales) and from
the pressureless dark matter distribution, imposing different
temperature-density relations on the evolved particle distributions. The
dark matter spectra are similar but not identical to those created from
the full gas distributions, showing that thermal broadening sets the
longitudinal (line-of-sight) scale of the \lya\ forest. The turnover
scales in the flux power spectrum and flux autocorrelation function are
determined mainly by thermal broadening rather than pressure. However,
the insensitivity to pressure arises partly from a cancellation effect
with a sloped temperature-density relation ($T \propto \rho^{0.6}$ in
our simulations): the high density peaks in the colder, lower pressure
simulation are less smoothed by pressure support than in the hotter
simulation, and it is this higher density gas that experiences the
strongest thermal broadening.  Changes in thermal broadening and
pressure support have comparably important effects on the flux
probability distribution (PDF), which responds directly to the gas
overdensity distribution rather than the scale on which it is
smooth. Tests on a lower resolution simulation ($2\times 144^3$ vs.\
$2\times 288^3$ particles in a $12.5\hmpc$ comoving box) show that our
statistical results are converged even at this lower resolution. While
thermal broadening generally dominates the longitudinal structure in the
\lya\ forest, we show in Paper~II that pressure support determines the
transverse coherence of the forest observed towards close quasar pairs.
\end{abstract}


\section{Introduction}\label{sec:intro}
The \lyman\ forest, caused by \lya\ absorption of neutral hydrogen atoms
along the line of sight to some distant source (usually a quasar), was
originally described in terms of discrete intervening gas ``clouds''
\citep{lynds71,sargent80}, analogous to clouds in the Galactic
interstellar medium.  In this picture, the velocity widths of the
observed absorption structures were primarily a consequence of thermal
motions of the absorbing atoms.  In the mid-1990s, three-dimensional
hydrodynamic simulations of cold dark matter (CDM) cosmological models
achieved remarkable success in reproducing the observed properties of
the \lya\ forest \citep{cen94,zhang95,hernquist96,miralda96,theuns98}.
In these simulations, most of the absorbing gas is at low density
($\rho/\bar{\rho}\sim0.1$--10) and naturally described as a continuously
fluctuating medium rather than a series of discrete structures
\citep{hernquist96,bi97,rauch97,croft97,croft98}.  The velocity width of
individual features is set largely by the Hubble flow across them,
reflecting the physical extent of the absorbing gas along the line of
sight \citep{hernquist96,weinberg97}.  The temperature of the
intergalactic medium (IGM), therefore, affects the structure of the \lya\
forest in two ways: by smoothing absorption along the line of sight
through the thermal motions of atoms, and by smoothing the physical
distribution of the gas in three dimensions through pressure support.
In this study, we use two smoothed particle hydrodynamics (SPH)
simulations with different photoionization heating rates---and thus
different IGM temperatures and amounts of gas pressure support---to
disentangle the relative effects of pressure support and thermal
broadening in the \lya\ forest.

Both hydrodynamical simulations \citep{katz96,miralda96,theuns98} and
analytic arguments \citep{hui97b} suggest that low-density intergalactic
gas should have a power-law temperature-density relation, i.e.,
$T=T_0(1+\delta)^{\alpha}$, where $1+\delta\equiv\rho/\bar{\rho}$ is the
local gas overdensity.  Reasonable assumptions for heating and cooling
rates yield $T_0\sim 10^4$\,K and $\alpha\sim0.6$
\citep{hui97b,theuns98}.  Observations, however, imply that the
normalization $T_0$ could be nearly twice as high, and the slope could
be much shallower or even inverted
\citep{schaye99,ricotti00,mcdonald01,bolton08}.  Regardless of the
parameter values, because the optical depth to \lya\ absorption is
related to both the temperature and the density by a power law, the
existence of the temperature-density relation (also called the
``equation of state'') implies a tight relation between the observed
\lya\ absorption and the IGM gas density.  Since the universe has $\sim
5$ times as much mass in dark matter as in baryons, we can expect in
general for intergalactic gas to trace the underlying dark matter on
scales above the gas Jeans length \citep{schaye01}.  The \lya\ forest
therefore provides a powerful tool for tracing the dark matter power
spectrum in the quasi-linear regime---modulo the effects of peculiar
velocities, thermal broadening, and gas pressure
\citep{croft98,croft99,croft02,mcdonald00,mcdonald06,viel04,viel06a}.

Pressure should be important on scales at or below the Jeans length,
\begin{equation}\label{eqn:jeans}
\lambda_J = c_s\sqrt{\frac{\pi}{G\rho}} = \sigma_{\rm th}\sqrt{\frac{5\pi}{3G\rho}}, 
\end{equation}
where $c_s=\sqrt{[5kT]/[3m]}=\sigma_{\rm th}\sqrt{5/3}$ is the speed of
sound in an ideal gas expressed as a multiple of the 1-D thermal velocity
$\sigma_{\rm th}$ \citep{miralda96,schaye01,desjacques05}.  Here $m$ is
the average mass of the gas particle; in an ionized primordial mixture
of hydrogen and helium, $m=0.59m_p$, where $m_p$ is the proton mass.
The density $\rho$ is the density of the gravitating medium, which is
dominated by dark matter, so we take $\rho =
\Omega_{m,0}\rho_{c,0}(1+z)^3 (1+\delta)$.  In comoving coordinates,
\begin{eqnarray}\label{eqn:jeanscomove}
\lambda_{J,{\rm comv}} & = & (1+z)\sigma_{\rm th}H_0^{-1}\sqrt{\frac{5\pi}{3}}\left[\frac{3}{8\pi}\Omega_{m,0}(1+z)^3(1+\delta)\right]^{-1/2}\nonumber\\
 &= & 782\,h^{-1}\,\mbox{kpc}\\
 &\times& \left(\frac{\sigma_{\rm
 th}}{11.8\,\rm{km}\,\rm{s}^{-1}}\right)\left[\left(\frac{\Omega_{m,0}(1+\delta)}{0.25\times(1+0)}\right)\left(\frac{1+z}{1+3}\right)\right]^{-1/2}, \nonumber
\end{eqnarray}
where we have normalized the thermal broadening velocity $\sigma_{\rm
th}$ to correspond to a fiducial temperature of $10^4$\,K.  By defining
a ``Jeans velocity'' as $v_J\equiv H\lambda_J$ we can find the relative
importance of the Jeans scale and $\sigma_{\rm th}$,
\begin{eqnarray}\label{eqn:vjsigth}
\frac{v_J}{\sigma_{\rm th}} &=&
\frac{2\pi\sqrt{10}}{3}(1+\delta)^{-1/2}\left[\frac{\Omega_{m,0}(1+z)^3 +
    \Omega_{\Lambda}}{\Omega_{m,0}(1+z)^3}\right]^{1/2}\\\nonumber
&\approx& 6.62(1+\delta)^{-1/2},
\end{eqnarray} 
which is essentially redshift-independent for the redshifts relevant to
the \lya\ forest.  The Jeans length of equation~(\ref{eqn:jeans})
divides stable from unstable modes in a static, homogeneous,
self-gravitating medium.  The IGM is expanding, inhomogeneous,
non--self-gravitating (because dark matter dominates), and evolving in
density and temperature on the same timescale that fluctuations grow.
Even for linear perturbations in a baryonic universe, the ``filtering
scale'' below which fluctuations growth is suppressed depends on the
thermal history of the gas rather than the instantaneous
temperature-density relation \citep{gnedin98}.  We therefore expect
equation~(\ref{eqn:jeans}) to describe the scale of gas pressure support
only at an order-of-magnitude level.  Equation~(\ref{eqn:vjsigth}) shows
that Jeans velocities and thermal velocities should be comparable at the
overdensities of typical \lya\ forest features, but the calculation is
not definitive enough to show whether one will dominate in practice.

Hence, to predict the statistical properties of the \lya\
forest in a given cosmological model, one must calculate the predicted
gas distribution.  The most reliable way to do this uses full
$N$-body plus hydrodynamic simulations, which include the gravity of
dark matter and gas and the additional effects of pressure, adiabatic
heating and cooling, heating by photoionization and shocks, and
radiative cooling.  However, large volume hydrodynamic simulations with
the necessary resolution are computationally intensive.
\citet{weinberg97} and \citet{croft98} show that one can achieve
reasonable accuracy in the \lya\ forest regime from pure dark matter
simulations, applying the temperature-density relation to the evolved
dark matter distribution to compute spectra.  Indeed, the log-normal
model \citep{bi97}, in which the dark matter distribution is computed
from the linear density field by an exponential transformation
\citep{coles91}, provides a qualitatively accurate physical model of the
\lya\ forest, sufficient for creating artificial spectra with reasonable
statistical properties.

A fast way to include the effects of gas pressure in an approximate way
is the hydrodynamic particle-mesh (HPM) method
\citep{gnedin98,ricotti00,meiskin01}.  HPM assumes that all the gas
follows the power-law IGM equation of state, and it uses a modification
of the standard (fast) particle-mesh $N$-body method to compute the sum
of gravitational forces and pressure gradient forces given this equation
of state.  In a detailed comparison of fully hydrodynamic SPH
simulations and the approximated HPM simulations, both evolved using
{\sc Gadget-2}, \citet{viel04} found that while the HPM approach does
converge to the SPH results, for some \lya\ forest properties, such as
the flux probability distribution or small-scale power spectrum, it can
differ from the SPH calculation by as much as 50\% at $z\sim 2$.  A
simpler alternative to HPM is to use a pure $N$-body simulation but
smooth the evolved dark matter distribution on the Jeans scale before
extracting \lya\ forest spectra \citep{zaldarriaga01,desjacques05}.
Because the Jeans smoothing is three-dimensional, it is not degenerate
with line-of-sight thermal broadening. \citet{zaldarriaga01} found that,
even though the thermal broadening dominates the pressure correction,
the value of the Jeans length becomes a large source of uncertainty in
cosmological inferences from the \lya\ forest if one tries to estimate
it from the data rather than predict it from theory.  Because the Jeans
length is expected to vary with the gas temperature and density, a
Zel'dovich-like scheme can be used when smoothing the dark matter
distribution in order to model these effects \citep{viel02a}.  Though we
do not carry out a comprehensive comparison of these methods here, we do
investigate the impact of pressure in detail, as well as compare full
SPH simulations to results derived from the dark matter distribution
alone.

The primary purpose of this paper and its companion (Peeples et al.\
2009, hereafter Paper~II) is to disentangle the roles of pressure
support and thermal broadening in the \lya\ forest by studying two SPH
simulations with different thermal histories and temperature-density
relations.  In addition to examining the physics of the \lya\ forest,
this study is motivated by observational evidence (discussed in
\S\,\ref{sec:rhot}) that the temperature of the IGM at $z\sim 2$--4 is
higher than expected from simple photoionization models by a factor of
1.5--2.  This evidence comes from analyses of data along single lines of
sight.  Because the higher pressure associated with hotter gas would
smooth the IGM in three dimensions, using closely paired lines of sight
to probe this coherence scale has been proposed as an alternative route
for inferring the temperature-density relation (J.\ Hennawi, private
communication, 2007).  However, before we can understand the relative
roles of temperature and pressure on the transverse structure of the
\lyman\ forest, we must first understand their longitudinal effects
along independent sightlines, which is the goal of this paper.  While
here we find that thermal broadening dominates pressure support in
setting the level of longitudinal structure in the \lya\ forest, in
Paper II we show that the gas Jeans length dominates the level of
coherence transverse to the line of sight.

This paper is organized as follows.  In \S\,\ref{sec:sims}, we describe
the SPH simulations we have evolved to investigate these effects.  In
\S\,\ref{sec:physics}, we describe the physics of the \lya\ forest in
these simulations and examine the impact of the thermal history on
observable spectra; this section also serves to review the physical
understanding of the high-redshift \lya\ forest that has emerged from
simulations and associated analytic work since the mid-1990s. We then
study these effects on several typical statistical measures for learning
about the IGM from the \lya\ forest in \S\,\ref{sec:stats}, with our
conclusions in \S\,\ref{sec:conc}.

\section{Simulations}\label{sec:sims}
We analyze two SPH simulations with identical initial conditions, one
with a fiducial photoionization heating rate and one with a heating rate
from photoionization that is four times higher than the fiducial;
hereafter, we refer to these simulations as the ``fiducial'' and ``H4''
simulations, respectively.  In the terminology of \citet{katz96}, we
compute the photoionization rates $\Gamma$ and photoionization heating
rates $\epsilon$ for the fiducial simulation assuming the
\citet{haardt01} quasar\,$+$\,galaxy photoionizing background, and for
the H4 simulation we increase $\epsilon_{\rm H\,I}$,
$\epsilon_{\rm He\,I}$, and $\epsilon_{\rm He\,II}$ by a factor
of four.  In principle, these heating rates could arise from a much
harder UV background spectrum that yields more residual energy per
photo-electron.  However, we do not propose any specific model for these
heating rates---they are a computationally simple way to obtain IGM
temperatures that are higher than those in the fiducial model and closer
to those estimated from observations (see \S\,\ref{sec:rhot} for further
discussion).

We evolve these SPH simulations using the parallel {\sc Gadget-2} code
\citep{springel05} to trace the evolution of $288^3$ dark matter and
$288^3$ gas particles in a $12.5\,h^{-1}$\,Mpc comoving cubic volume
from $z=15$ to $z=2$.  We also evolved another simulation using the
fiducial heating rates and the same initial conditions but only
$2\times144^3$ particles to serve as a test for resolution convergence.
We adopt a standard $\Lambda$CDM cosmological model with the parameters
$(\Omega_M,\Omega_{\Lambda},\Omega_b,h,\sigma_8,n_s) =
(0.25,0.75,0.044,0.7,0.8,0.95)$, all of which are in good agreement with
the {\em Wilkinson Microwave Anisotropy Probe} (WMAP) five-year results
\citep{hinshaw09}; our choice of $\sigma_8=0.8$, however, is somewhat
lower than what is typically considered for \lya\ studies.  These
parameters give a mass per SPH particle of $1.426\times10^6\msun$, which
is much less than the typical Jeans mass of
$M_J\equiv\rho\lambda_J^3\sim7\times10^9\msun$.  The spline
gravitational force softening has an equivalent Plummer length of
$0.875\,h^{-1}$\,kpc comoving ($\sim 1/50$ of the initial particle grid
spacing).  The SPH smoothing lengths are chosen to enclose $33\pm 2$
neighbors within the smoothing kernel.  The simulation incorporates
standard heating and atomic cooling processes and the standard {\sc
Gadget-2} treatment of star formation and metal enrichment.  These
simulations do not incorporate galactic winds, but these should have
very little effect on the \lya\ forest
\citep{kollmeier06,bertone06,marble08b}.

We extract spectra from the SPH gas distribution using
TIPSY\footnote{University of Washington version}, as described by
\citet{hernquist96}.  Following common practice, we rescale the
intensity of the UV background so that the mean flux decrement of the
extracted spectra matches observations (Table~\ref{tab:rhot}), as the
mean decrement itself is much better known than the background intensity
(see discussions by, e.g., \citealt{croft02,marble08b}).  Our extracted
spectra have 1250 pixels across the $12.5h^{-1}$\,Mpc volume, making the
pixel size $\approx 1$\kms, which is well below the smoothing scale
imposed by thermal broadening.

\section{Physics of the Lyman-$\alpha$ forest}\label{sec:physics}

\subsection{The Temperature-Density Relation}\label{sec:rhot}
In the absence of shock heating, the evolution of the temperature of the
IGM is described by
%\begin{eqnarray}\label{eqn:dTdz}
\begin{equation}\label{eqn:dTdz}
\frac{\mbox{d}T}{\mbox{d}z}  =  \frac{2T}{1+z} +
       \left[\frac{2T}{3(1+\delta)}\right]\frac{\mbox{d}\delta}{\mbox{d}z} 
     - \left[\frac{T}{m}\right]\frac{\mbox{d}m}{\mbox{d}z} + \frac{2}{3k_{\mbox{\tiny $B$}}n_b}\frac{\mbox{d}Q}{\mbox{d}z},
\end{equation}
%\end{eqnarray}
as shown in detail by \citet{hui97b}.  Here, $\mbox{d}\delta/\mbox{d}z$
and $H(z)$ depend on the cosmology, the overdensity $1+\delta\equiv
(\rho_{\rm gas}/\bar{\rho}_{\rm b})$, $m$ is the mean particle mass, and
$\mbox{d}Q/\mbox{d}z$ is the net power per unit volume owing to the
ambient radiation field.  The first two terms in
equation~(\ref{eqn:dTdz}) describe heating and cooling owing to
adiabatic processes.  After reionization, the change in temperature
owing to the change in the ionization fraction (the third term in
equation~[\ref{eqn:dTdz}]) is effectively zero at all redshifts and
relevant densities.

\begin{figure}
\includegraphics[width=0.48\textwidth]{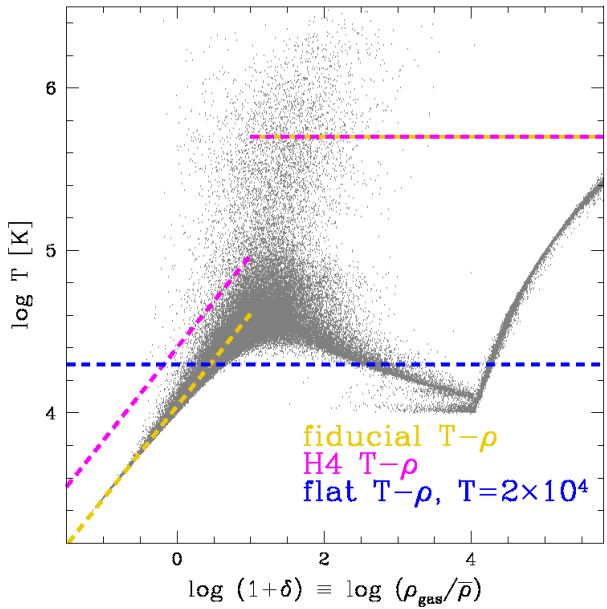}
\caption{\label{fig:rhot}Distribution of 1\% of the gas particles for
  the fiducial simulation in the temperature-density plane at $z=3$.  In
  subsequent analyses, we use either the simulation temperatures
  themselves or one of the three plotted \rhot\ relations ({\em dashed
  lines}).}
\end{figure}

Hydrogen reionization produces one energetic photoelectron per hydrogen
atom, and it is expected to heat the IGM to a temperature $T\sim
2$--$5\times 10^4$\,K, depending on the spectral shape of the ionizing
sources and radiative transfer effects \citep{miralda94}.  Thereafter,
adiabatic cooling reduces the overall temperature, but denser regions
remain hotter because they have higher neutral fractions and thus higher
photoionization heating rates.  At a given redshift, the simulated
temperature-density (\rhot) relation of the photoionized medium can be
well approximated by a powerlaw,
\begin{equation}\label{eqn:rhot}
T(z) = T_0(z) (1+\delta)^{\alpha(z)}.
\end{equation}
The slope $\alpha(z)$ approaches $0.6$ well after reionization.
Evolution with the \citet{haardt01} UV background spectrum and $\Omega_b
h^2\approx 0.022$ yields $T_0\approx10^4$\,K at $z=2$--4, for
reionization at $z\gtrsim7$ (e.g., \citealt{hui97b,theuns98,dave99};
Table~\ref{tab:rhot}).  Figure~\ref{fig:rhot} shows the distribution of
SPH particles for our fiducial simulation in the temperature-overdensity
plane at $z=3$.  At low overdensities ($1+\delta \lesssim 10$), most of
the gas falls along a tight locus, as expected from the above
discussion.  At higher densities ($1+\delta \gtrsim 100$), the gas has
begun to cool to form galaxies.  The higher temperature gas has been
shock heated.  (The apparent increase in temperature at $1+\delta \sim
10^4$ owes to the way {\sc Gadget-2} treats the multiphase interstellar
medium.)  For many of our subsequent analyses, we will isolate physical
effects by imposing one of the three \rhot\ relations, denoted by dashed
lines in Figure~\ref{fig:rhot}.  Specifically, we assign each gas (or
dark matter) particle the temperature implied by its overdensity and a
given \rhot\ relation before extracting spectra.  The yellow line is an
eyeball fit to the \rhot\ relation in the fiducial simulation, while the
pink line is the corresponding fit to the H4 relation; the H4 \rhot\
normalization is $\sim 2.3$ times higher than the fiducial
normalization.  In both of these cases we set gas with $1+\delta>10$ to
a ``shocked'' temperature of $T=5\times10^5$\,K, so that this high
density gas will be entirely ionized and thus not contribute to the
\lya\ forest.  The mass- and volume-fraction of gas in these high
density regions is relatively tiny; for all of the statistics presented
here, the fiducial and H4 gas distributions using the imposed fiducial
and H4 \rhot\ relations, respectively, yield nearly identical results to
those obtained using the actual temperatures calculated by {\sc
Gadget-2}.  Parameters for these fits are listed in
Table~\ref{tab:rhot}.  For some comparisons, we impose a flat \rhot\
relation, with all particles set to $T=2\times10^4$\,K, as shown by the
blue dashed line.

\begin{figure*}
\begin{center}
\includegraphics[height=0.9\textwidth,angle=270]{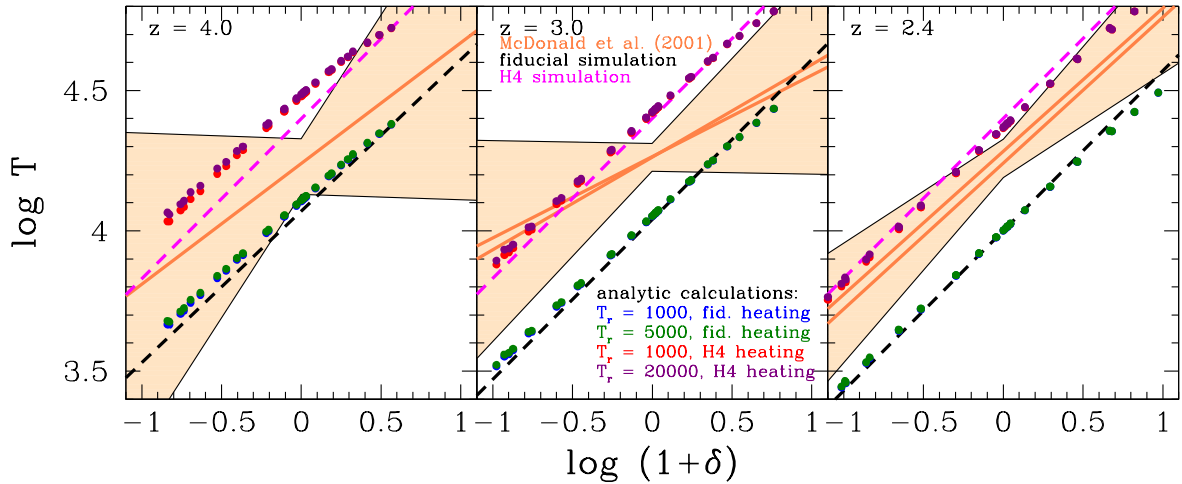}
\end{center}
\caption{\label{fig:mcd01}A comparison of the temperature-density
  relation evolution from simulations ({\em dashed lines}), calculations
  using equation~(\ref{eqn:dTdz}) ({\em points}), and observations ({\em
  shaded regions}).  The shaded region corresponds to the 1-$\sigma$
  error region from \citet{mcdonald01}, with the orange lines indicating the
  best-fit relations at each redshift.  (There are two such lines for
  $z=3$ and 2.4 because \citeauthor{mcdonald01}\ used simulations at two
  different redshifts to compare to these observations.)  At $z=3$, the
  dashed pink and black lines are the same as the dashed pink and gold
  lines, respectively, in Fig.\,\ref{fig:rhot}.}
\end{figure*}

Although most \lya\ forest features are broadened by Hubble flow, the
narrowest features arise at velocity caustics and have widths set by
thermal broadening.  Analyses of observed line width distributions imply
$T_0\approx1.5$--$2\times 10^4$\,K at $2\leq z\leq 4$
(\citealt{gnedin98,ricotti00,schaye00,mcdonald01}; we adopt
\citeauthor{mcdonald01}'s constraints in Table~\ref{tab:rhot} and
Figure~\ref{fig:mcd01}).  This is significantly hotter than the value
$T_0\approx 10^4$\,K expected for a \citet{haardt01} ionizing
background.  \citet{theuns00} and \citet{zaldarriaga01} find a similar
result by fitting the small-scale cutoff of the one-dimensional flux
power spectrum.  More recently, \citet{bolton08} have fit the flux
probability distribution function (PDF) in high-resolution spectra
inferring a similar, high $T_0$ and a shallow, possibly inverted
($\alpha < 0$) slope. \citet{lidz09} find $T_0 \approx 2\times 10^4$\,K
over the redshift range $z=2$--4, using a wavelet analysis of 40
high-resolution spectra. In all cases, the observations are interpreted
by comparing them to a suite of cosmological simulations, which
incorporate many parameters in addition to the \rhot\ relation itself.

According to equation~(\ref{eqn:dTdz}), there are two basic ways to
increase the gas temperature: either have hotter ``initial conditions,''
or a higher heating rate $\mbox{d}Q/\mbox{d}z$.  Even with the highest
plausible reionization temperatures, it is difficult to reproduce the
inferred $T_0$ at $z=3$, given the observational evidence that $z_{\rm
reion} \ge 7$ (\citealt{fan06,hinshaw09}; see also \citealt{hui03}).
Energy injection from \ionm{He}{II} reionization over an extended period
of time could also keep the IGM hot down to $z\sim 3$
\citep{bolton08,furlanetto08}, though even this mechanism appears unable
to produce an inverted \rhot\ relation \citep{mcquinn09}.  A harder
photoionizing spectrum produces more energy input per photoelectron and
thus a higher IGM temperature.\footnote{Note that in photoionization
equilibrium, higher intensity background radiation---with the same
spectrum---affects only the photoionization and recombination rates,
{\em not} the electron temperature.}  It is this solution that we adopt
for our H4 simulation, though the heating rates we adopt correspond to
an implausibly hard spectrum.  The structure of the \lya\ forest should
depend on the \rhot\ relation but be fairly insensitive to the detailed
mechanism that produces it, so we expect our conclusions about the
impact of pressure support to apply to a broad range of such mechanisms.
However, the tension between the \rhot\ relations predicted from theory
and those inferred from observations remain puzzling, and it is not
clear whether the resolution lies in modest changes (e.g., temperatures
at the low end of observational estimates and a spectrum that is harder
than conventionally assumed) or a physical process that has not yet been
identified.  Precisely because of this tension, it is important to
understand the physics and observational consequences of pressure
support in the \lya\ forest, as we study in this paper and in Paper~II.

In Figure~\ref{fig:mcd01}, the dashed lines show the approximate \rhot\
relationships found in our two SPH simulations; the $z=3$ dashed lines
are the same as the low-density yellow and pink lines in
Figure~\ref{fig:rhot}.  The temperature-density parameters for the
observations and our simulations at each redshift are given in
Table~\ref{tab:rhot}.  The fiducial simulation clearly disagrees with
the observations at $z=3$ and 2.4, while the H4 simulation is in closer
agreement but somewhat too hot, i.e., the two simulations bracket the
central observation estimates. The points in Figure~\ref{fig:mcd01} are
numerical integrations of equation~(\ref{eqn:dTdz}), where we
approximate the overdensity evolution using a modified Zel'dovich
approximation \citep{reisenegger95}. \ionm{H}{I} reionization is modeled
by initializing the temperature at a ``reionization temperature'' $T_r$
for all overdensities at a reionization redshift $z_{\rm reion}=9.45$;
the temperature-density relation is fairly independent of $T_r$ by
$z=4$, assuming reionization occurs at $z\gtrsim 6$.

\begin{table*}
 \centering
 \caption{\label{tab:rhot}Observed mean flux decrements $\langle D \rangle\equiv
    \langle 1-e^{-\tau}\rangle$ are from \citet{mcdonald00} and observed
    temperature-density relations ($T = T_0[1+\delta]^{\alpha}$) are
    from \citet{mcdonald01}.}
 \begin{tabular}{ccrccccc}
\hline\hline $z$ & $\langle D\rangle$ & observed $T_0$ [K] & observed $\alpha$ &
   fiducial $T_0$ [K] & fiducial $\alpha$ & H4 $T_0$ [K] & H4
       $\alpha$\\\hline
4.0 & $0.525\pm 0.012$ &   $17400\pm 3900$ & $0.43\pm 0.45$ & $11700$ & $0.54$ & 28200 & 0.55 \\\hline
3.0 & $0.316\pm 0.023$ &  $18300\pm 1800$ & $0.33\pm 0.26$ & $11000$ & $0.57$ & 25000 & 0.57\\
    &  & or $18400\pm 2100$ & $0.29\pm 0.30$ & & & & \\\hline
2.4 & $0.182\pm 0.021$ &    $17400\pm 1900$ & $0.52\pm 0.14$ & $10000$ & $0.56$ & 23000 & 0.57\\
    &  & or $19200\pm 2000$ & $0.51\pm 0.14$ & & & & \\\hline
\end{tabular}
\end{table*}

\subsection{The Fluctuating Gunn-Peterson Approximation}\label{sec:fgpa}
The \ionm{H}{I} optical depth $\tau_{\mbox{\tiny HI}}$ is proportional to
the neutral hydrogen density, with
\begin{equation}\label{eqn:tauGP}
\tau_{\mbox{\tiny HI}} = \frac{\pi e^2}{m_e c}f\lambda_0 H^{-1}(z)n_{\mbox{\tiny HI}},
\end{equation}
where $f_{\lambda}$ is the oscillator strength of the \lya\ transition
and $\lambda_0$ is the center of the \lya\ transition, 1216\AA\
\citep{gunn65,miralda93}.  In the limit of high ionization fraction, the
neutral hydrogen density,
\begin{equation}\label{eqn:hi}
n_{\mbox{\tiny HI}} = \left(\frac{\alpha_{\mbox{\tiny HII}}}{\Gamma_{\mbox{\tiny UV}}}\right)n_e n_{\mbox{\tiny H}},
\end{equation} 
is proportional to the gas density squared; $\Gamma_{\mbox{\tiny UV}}$
is the photoionization rate owing to the ambient UV background.  At the
relevant temperatures ($T \ll 10^6$), the recombination coefficient
$\alpha_{\mbox{\tiny HII}}$ is proportional to $T^{-0.7}$
\citep{katz96}.  Following \citet{rauch97} and \citet{croft98}, we can
combine equations~(\ref{eqn:tauGP}) and (\ref{eqn:hi}) with a power-law
\rhot\ relation to obtain the optical depth
\begin{equation}\label{eqn:tau}
\begin{array}{l}\displaystyle
\tau_{\mbox{\tiny HI}} = 1.54\times 
       \left(\frac{T_0}{10^4\,\mbox{K}}\right)^{-0.7}
       \left(\frac{10^{-12}\,\mbox{s}^{-1}}{\Gamma_{\mbox{\tiny UV}}}\right)
       \left(\frac{1+z}{1+3}\right)^6 \left(\frac{0.7}{h}\right)\hspace*{-0.76em}\vspace*{0.6em}\\\displaystyle
\times \left(\frac{\Omega_{b,0}h^2}{0.02156}\right)^2 
       \left[\frac{4.0927}{H(z)/H_0}\right]
       \left(1+\delta\right)^{2-0.7\alpha} 
       \left[1+\frac{1}{H(z)}\frac{{\rm d}V_{\rm los}}{{\rm d}x}\right]^{-1}\nonumber.\hspace*{-0.76em}
\end{array}
\end{equation}  % \left[\frac{0.44}{X/(X+4Y)}\right] \\
The last factor represents the impact of line-of-sight peculiar velocity
gradients ${\rm d}V_{\rm los}/{\rm d}x$, which change the density of
atoms in frequency space relative to real space.

Independently of the \lya\ forest, the photoionization rate
$\Gamma_{\mbox{\tiny UV}}$ is difficult to constrain; the other
quantities entering the normalization pre-factor of
equation~(\ref{eqn:tau}) are also uncertain.  We therefore follow
standard practice \citep[e.g.,][]{miralda96,marble08b} and choose
$\Gamma_{\mbox{\tiny UV}}$ for each model so that it reproduces the
observed mean flux decrement $\langle D\rangle\equiv\langle 1-F\rangle =
\langle 1-\exp(-\tau_{\mbox{\tiny HI}})\rangle$ at the redshift under
investigation.  Specifically, we adopt $\langle D\rangle$ values from
\citet{mcdonald00}, which are listed in Table~\ref{tab:rhot}; at the
overlapping redshifts ($z=2.4,\,3$) these $\langle D\rangle$ are
consistent with the more recent measurements of \citet{kim07}.  If
$\Gamma_{\mbox{\tiny UV}}$ were perfectly known, then $\langle D\rangle$
could itself be used as a diagnostic of the IGM temperature $T_0$, but
in practice it is not well enough known.  Hence, we rely on structure in
the \lya\ forest for IGM diagnostics and cosmological tests.

Our results below rely on our full hydrodynamical simulations, but
equation~(\ref{eqn:tau}) is useful to understand our results, and it can
be a useful basis for simpler analytic or numerical treatments.  It is
often referred to as the ``fluctuating Gunn-Peterson approximation''
\citep[FGPA;][]{weinberg97,croft98} because it describes \lya\
absorption as a continuous phenomenon analogous to the Gunn-Peterson
(\citeyear{gunn65}) effect, but arising in a fluctuating medium.  As
written, it ignores thermal broadening (a temperature- and therefore
density-dependent convolution) and shock heating (i.e., gas not falling
on the temperature-density relation).  Also, while $1+\delta\equiv
\rho_{\mbox{\tiny gas}}/\bar{\rho}_{\mbox{\tiny gas}}$, the gas
overdensity is often approximated as the dark matter overdensity
$\rho_{\mbox{\tiny DM}}/\bar{\rho}_{\mbox{\tiny DM}}$ in $N$-body
simulations, perhaps smoothed by an effective Jeans length.

\subsection{The \lya\ forest in the fiducial case}\label{sec:fid}
\begin{figure*}
\includegraphics[width=\textwidth]{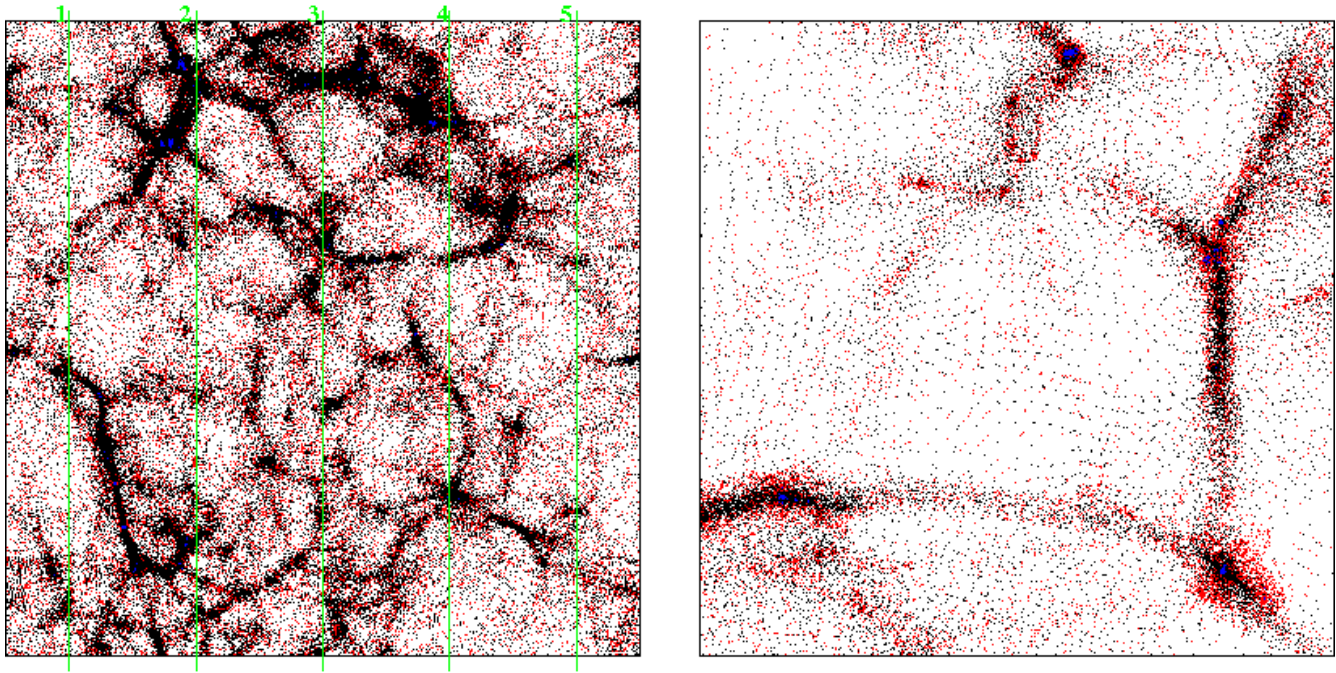}
\caption{\label{fig:slices}A $125\,h^{-1}$\,kpc thick slice ({\em left})
  and a $125\times125\,h^{-1}$\,kpc comoving region ({\em right}) of the
  fiducial simulation at $z=3$ with gas particles ({\em red}), dark
  matter particles ({\em black}), and star particles ({\em blue}) all
  shown.  The five green lines ({\em left}) denote sightlines referred
  to in subsequent figures.}
\end{figure*}

Figure~\ref{fig:slices} shows, on the left, a slice through our fiducial
simulation at $z=3$, with a depth of $125\,h^{-1}$\,kpc.  The
small-scale filamentary structure of the high redshift universe is
evident. While the dark matter and baryons have similar large-scale
structure, the zoom-in panel on the right shows that the gas
distribution is more diffuse.  In particular, the densest filaments of
dark matter lie within thicker filaments of gas.  This difference
reflects the impact of gas pressure support.

\begin{figure*}
\includegraphics[height=0.9\textheight]{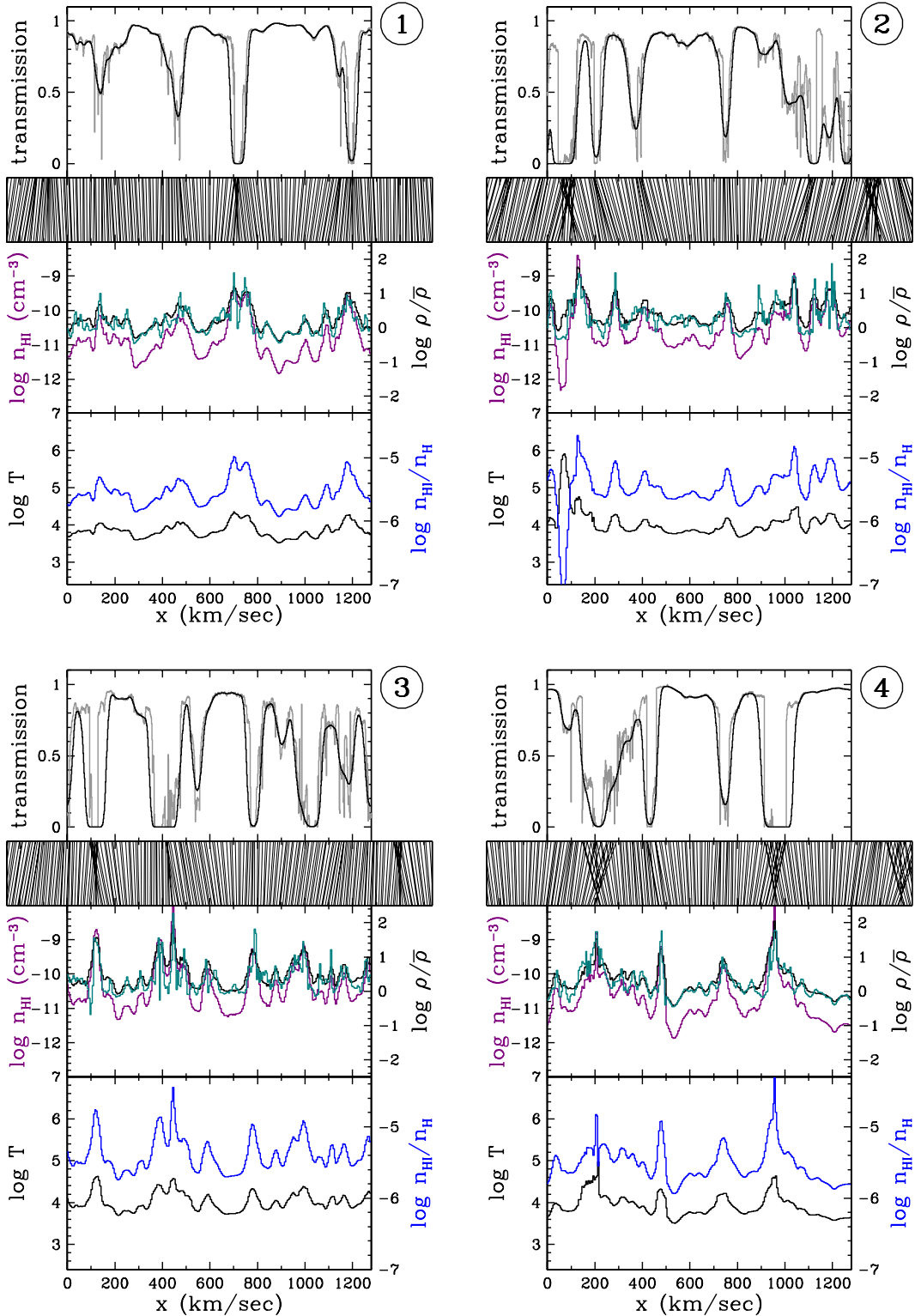}
\caption{\label{fig:zoo}Physical quantities and \lya\ forest spectra
  along four of the sightlines labelled in Figure~\ref{fig:slices};
  velocity $v=0$ corresponds to the top of
  Figure~\ref{fig:slices}. Shown in real space are the temperature ({\em
  black, bottom}), the neutral hydrogen fraction ({\em blue, bottom}),
  the \ionm{H}{I} number density ({\em purple, middle}), the gas
  overdensity ({\em black, middle}), and the dark matter overdensity
  ({\em cyan, middle}).  The bars separating the top and middle panels
  show the effects of peculiar velocities when transitioning from real
  space to the observed flux transmission with ({\em top, black}) and
  without ({\em top, grey}) thermal broadening.  See \S\,\ref{sec:fid}
  for more details.  }
\end{figure*}

Figure~\ref{fig:zoo} shows how density, temperature, velocity, and
thermal broadening combine to produce \lya\ forest spectra, along the
sightlines marked 1--4 in Figure~\ref{fig:slices}.  In each panel, the
bottommost plot shows the gas temperature in black and the neutral
hydrogen fraction in blue.  Because $n_{\mbox{\tiny HI}}/n_{\mbox{\tiny
H}}\propto n_{\mbox{\tiny H}}T^{-0.7}$ and $T\propto n_{\mbox{\tiny
H}}^{0.6}$ for gas on the \rhot\ relation, the neutral fraction and
temperature are positively correlated at most temperatures.  However,
high temperature gas has usually been shock heated off the \rhot\
relation, and the recombination coefficient $\alpha_{\mbox{\tiny HII}}$
falls more steeply than $T^{-0.7}$ at high temperature \citep{katz96},
causing the neutral fraction to decrease dramatically (see, e.g., the
feature near 100\,km\,s$^{-1}$ in sightline \#2).  In the middle graphs,
we plot the neutral hydrogen number density [cm$^{-3}$] in purple, using
the scale on the left-hand axis.  The gas and dark matter overdensities
are plotted in black and cyan, respectively, using the scale on the
right-hand axis.  In general, the dark matter and gas overdensities are
similar, but the dark matter has sharper, higher overdensity peaks as
was seen visually in Figure~\ref{fig:slices}.  As expected from the
\rhot\ relation, the gas overdensity and the gas temperature follow one
another except at very high gas overdensity. The highest gas
overdensities correspond to condensed halos---i.e., galaxies---and thus
the gas has cooled to lower temperatures in these regions.  The neutral
hydrogen density shows more variation than the gas density because it is
proportional $n_{\mbox{\tiny H}}^2$ (actually $\rho^{1.6}$ once
temperature effects are included).  The topmost plot has the transmitted
flux $F\equiv\exp(-\tau_{\mbox{\tiny HI}})$ in black and the transmitted
flux that would be observed in the absence of thermal broadening in
grey.  The lines between the middle plot and the top plot show the
effects of peculiar velocities when converting from neutral hydrogen
density in physical space to an observable flux in velocity
space.\footnote{The bar-like appearance of the lines showing the effects
of peculiar velocity also lends these complicated plots the name of
``zoo plots.''}  Features contracting along the line of sight cause
physically distinct gas regions to converge to the same region of the
observed spectrum.  However, if the connecting lines do not cross, then
the region still has net expansion.  Both the thermally broadened and
non-thermally broadened spectra have the same characteristic broad
features, implying that residual Hubble flow dominates the velocity
width of these features.  However, thermal broadening smooths the small
scale roughness.  At velocity caustics (converging lines in
Figure~\ref{fig:zoo}), peculiar velocities cancel the Hubble flow, and
the features do become narrower when one removes the thermal broadening.

\subsection{Impacts of temperature and pressure on gas evolution}\label{sec:compare}
\begin{figure}
\begin{center}
\includegraphics[width=0.48\textwidth]{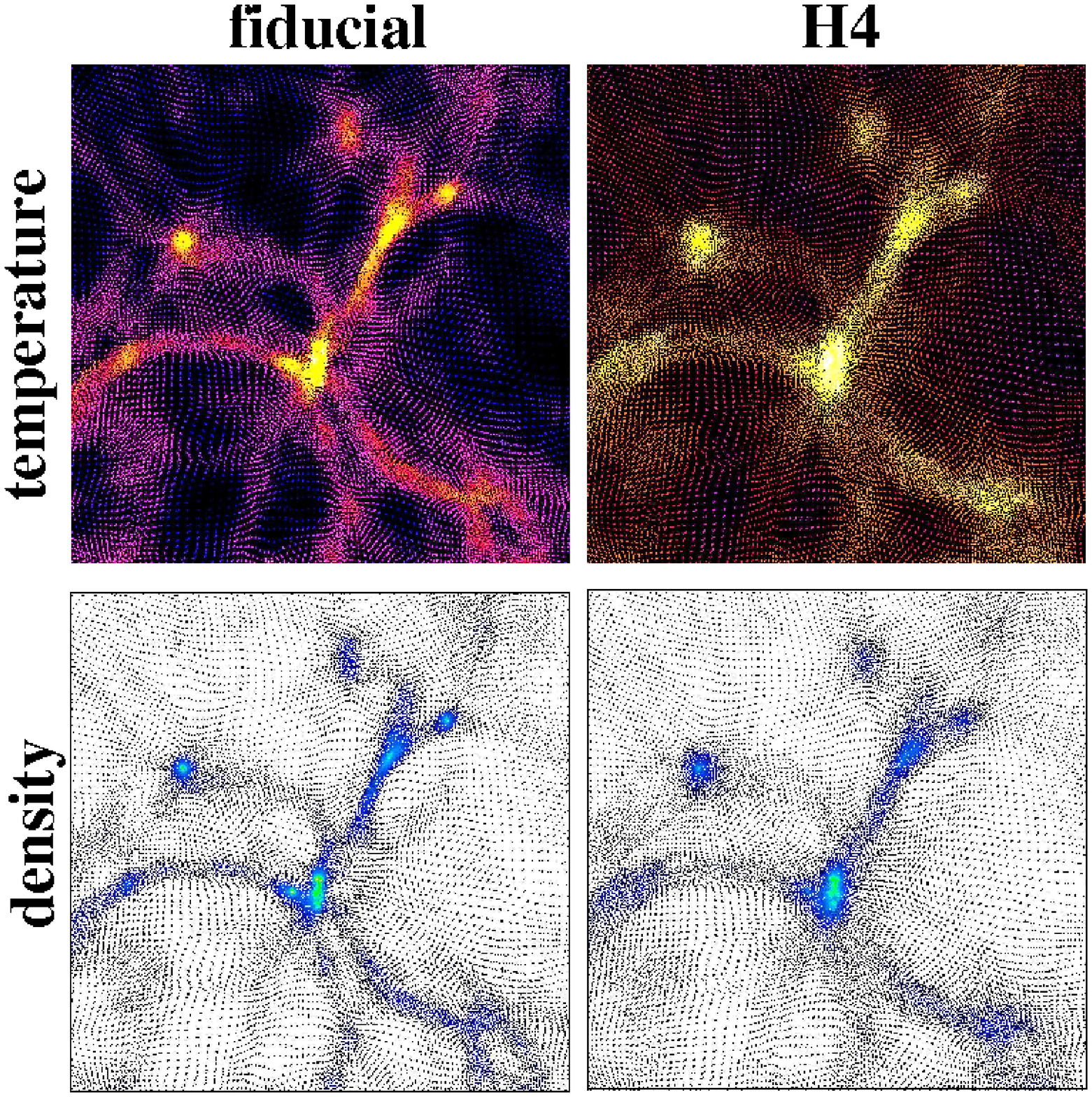}
\end{center}
\caption{\label{fig:temprhoevol} Temperature ({\em top}) and density
  ({\em bottom}) evolution for gas particles in a $2.5\times 2.5\times
  0.05$\,$h^{-1}$\,Mpc comoving slice at $z=3$ in the fiducial ({\em
  left}) and H4 ({\em right}) simulations. The temperature scale runs
  logarithmically from from $\log T = 3.5$ ({\em black/dark blue}) to
  $\log T = 5.5$ ({\em yellow/white}). The density scale runs from $\log
  (\rho_{\mbox{\scriptsize gas}}/\bar{\rho}_{\mbox{\scriptsize gas}})
  \equiv \log(1+\delta) = -1.3$ ({\em black}) to $\log (1+\delta) = 1.3$
  ({\em green}). }
\end{figure}

The obvious consequence of having a higher photoionization heating rate
is that gas temperatures in the H4 simulation are higher than in the
fiducial simulation, as shown at $z=3$ in the top panels of
Figure~\ref{fig:temprhoevol}.  A more subtle effect, is that the larger
Jeans length of the H4 simulation smooths the gas distribution, as shown
in the density-coded bottom panels of
Figure~\ref{fig:temprhoevol}. There is a relative paucity of very dense
clumps in the H4 simulation; this is especially noticeable in the lower
density filaments.

\begin{figure}
\includegraphics[height=0.48\textwidth,angle=270]{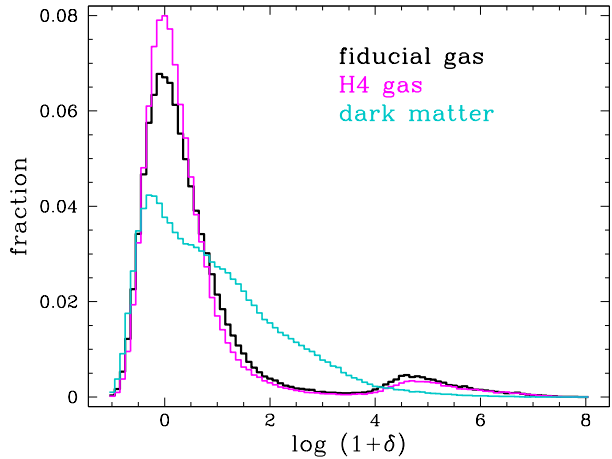}
\caption{\label{fig:rhohist}Distributions of gas overdensities at $z=3$
  in the fiducial ({\em black}) and H4 ({\em pink})
  simulations, as well as the dissipationless dark matter ({\em cyan})
  in the fiducial simulation.}
\end{figure}

Figure~\ref{fig:rhohist} plots the distribution of gas overdensity in
the fiducial and H4 simulations.  The visual differences in the right
panel of Figure~\ref{fig:temprhoevol} manifest themselves as a higher
frequency of particles with $\rho_{\rm gas}/\bar{\rho}_{\rm gas}\sim
10$--100 in the fiducial simulation and a higher frequency of $\rho_{\rm
gas}/\bar{\rho}_{\rm gas}\sim 1$ particles in the H4 simulation.  This
difference leads to a cancellation in many of the statistical
comparisons in \S\,\ref{sec:stats}; though the H4 gas is hotter (and
thus has more thermal broadening), its higher pressure implies that
there there is relatively less high-density and therefore relatively
higher-temperature gas.  For comparison, we also show in
Figure~\ref{fig:rhohist} the density distribution of the dark matter,
with densities computed using the SPH smoothing kernel.  Because the
dark matter is pressureless, it typically reaches higher overdensities,
though it does not achieve the highest overdensities seen in the gas
distributions because these arise from dissipation, i.e. cooling.
Extrapolating from the density distributions in our low-resolution
simulation, the dark matter density distribution would probably be
somewhat less skewed if we increased the mass resolution of the
simulation, but at low overdensity the gas distributions are essentially
converged since we resolve the Jeans mass.

\section{Effects of pressure and thermal broadening on the \lya\ forest}\label{sec:stats}
To isolate the effects of gas pressure and thermal broadening in our
subsequent analyses, we examine both the original gas particle
distributions and distributions with one of the three imposed \rhot\
relations illustrated in Figure~\ref{fig:rhot}.  For an imposed \rhot\
relation, we replace each gas particle's temperature by the temperature
that corresponds to its overdensity.  We use the same procedure to
create \lya\ forest spectra from the dark matter distribution
(``gasifying'' the dark matter).  The fiducial \rhot\ relation (yellow
in Figure~\ref{fig:rhot}) matches that found for the IGM in the fiducial
simulation; above $1+\delta=10$, the gas is ``shock heated'' to a
temperature of $5\times 10^5$\,K.  Likewise, the pink lines in
Figure~\ref{fig:rhot} show the \rhot\ relation used to mimic the H4
simulation.  Figure~\ref{fig:rhohist} shows that gas distributions with
different pressure also sample the \rhot\ relation differently, and if
the \rhot\ relation is sloped they will therefore experience different
thermal broadening For example, using the fiducial \rhot\ relation, the
H4 simulation would have less high density gas with large thermal
broadening, and the dark matter distribution would have more.  We,
therefore, consider an additional, flat \rhot\ relation, with $T=2\times
10^4\,$K at all densities, so that we can examine the effects of
pressure in the presence of pressure-independent thermal-broadening.

We now turn to the effects of pressure support and the \rhot\ relation
on the \lya\ forest spectra (\S\,\ref{sec:spectra}) and on flux
statistics (the flux power spectrum, \S\,\ref{sec:fluxpk}; the
autocorrelation function, \S\,\ref{sec:autocorr}; and the probability
distribution function, \S\,\ref{sec:pd}. For all these analyses we use
the same sightlines in each simulation to ensure the effects of sample
are variance the same.

\subsection{Spectra}\label{sec:spectra}
\begin{figure*}
\includegraphics[width=0.95\textwidth]{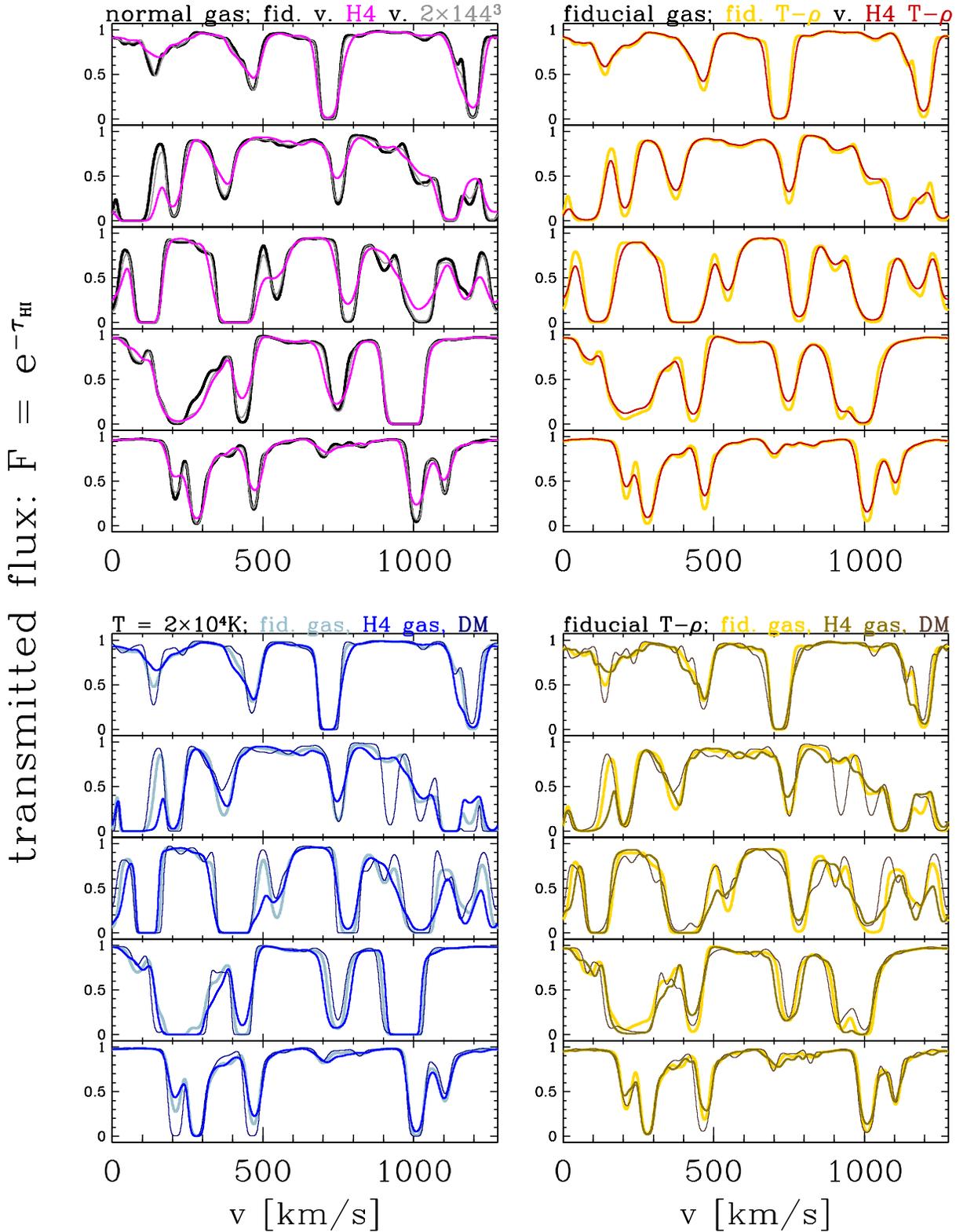}
\caption{\label{fig:spectra}Comparison of spectra along the sightlines
  labelled in Figure~\ref{fig:slices}, isolating different physical
  effects. Top left: full effects of thermal broadening, thermal history
  and resolution are shown.  Top right: effects of thermal broadening
  are isolated. Bottom left: effects of pressure support are isolated.
  Bottom right: effects of pressure support and different samplings of
  the underlying overdensity distribution are isolated.  See
  \S\,\ref{sec:spectra} for details.}
\end{figure*}

Figure~\ref{fig:spectra} shows how pressure and the \rhot\ relation
affect the \lya\ spectra along the five sightlines labelled in
Figure~\ref{fig:slices}, with sightline \#1 corresponding to the topmost
spectrum in each panel; as in Figure~\ref{fig:zoo}, $v=0$ corresponds to
the top of Figure~\ref{fig:slices}.  Any differences between the models are
usually most noticeable in unsaturated lines.  The top-left panel shows
spectra computed directly from the simulated gas distributions, with
differences in pressure support, differences in thermal broadening
because of the different \rhot\ normalization, and differences in thermal
broadening because of the sampling of the \rhot\ relation.  Light grey
lines show spectra from our lower resolution simulation.  The impact of
resolution is generally very small, but there are some slight
differences.

There are two notable differences between the fiducial and H4 spectra
exemplified by the $v\sim 450$\,km\,s$^{-1}$ feature in sightline \#5
and the $v\sim 750$\,km\,s$^{-1}$ feature in sightline \#4. First,
features in the H4 spectra are broader than in the fiducial spectra,
which could arise because of greater thermal broadening and/or because
of the larger Jeans scale (and thus the larger $H\lambda_J$).  Second,
features in the H4 spectra are not as deep as in the fiducial case
because the hotter gas does not reach as high overdensity and/or because
higher thermal broadening smears out inherently sharp features.

The top right panel isolates the impact of thermal broadening, applying
the fiducial and H4 \rhot\ relations to the gas distribution of the
fiducial simulation.  The lower-left panel isolates the impact of
pressure, applying the flat \rhot\ relation to the fiducial and H4 gas
distributions and to the dark matter distribution from the fiducial
simulation.  In some cases, such as the $v\sim 150$\kms\ feature in
sightline~\#1, the $v\sim 1025$\kms\ feature in sightline~\#3, and the
$v\sim 250$\kms\ feature in sightline~\#4, differences in the full
simulation spectra (upper left) are largely erased in the upper right
panel but remain similar in the lower left panel, which shows that they
mostly arise from different pressure effects in the two simulations
rather than from differences in the thermal broadening.  There are fewer
cases of the reverse, where differences present in the full spectra
remain in the upper right panels but disappear in the lower left, though
some are visible in sightline~\#5.  The differences in the upper-right
(thermal broadening isolated) are always systematic, with lower thermal
broadening yielding deeper and slightly narrower features.  The
differences between the H4 and fiducial gas distributions in the
lower-left (pressure isolated) are more random, and we will see below
that their statistical signature is weaker.

The pressureless dark matter distribution does lead to spectra with more
small scale structure, as shown by the thin dark lines in the lower
left.  However, despite the clear differences between the gas and dark
matter distributions evident in Figures~\ref{fig:slices}, \ref{fig:zoo},
and \ref{fig:rhohist}, the differences in the spectra created from these
distributions are small.  In the absence of thermal broadening, dark
matter spectra are much more jagged than their gas counterparts shown in
Figure~\ref{fig:zoo}, but realistic thermal broadening masks these
differences to a large extent.  The velocity widths of most features are
not sensitive to the level of thermal broadening (as seen in the upper
right) but neither are they set by the Jeans scale, or else the dark
matter and gas spectra would have greater differences.  Instead, thermal
broadening erases the finest scale structures in the density field so
that typical features correspond to coherent, moderate overdensity
structures expanding with the Hubble flow.

Adopting the fiducial \rhot\ relation in place of the flat \rhot\
relation (bottom right) makes only a small difference to the usual
appearance of the spectra (top left).  Features that are flat-bottomed in the dark
matter spectra with the flat \rhot\ relation often become less saturated
with the fiducial \rhot\ relation, such as the $v\sim 775$\kms\
feature in sightline~\#3 and the $v\sim 1000$\kms\ feature in
sightline~\#5.  In these regions, the dark matter overdensity is high,
so with $T\propto(1+\delta)^{0.6}$ they have higher thermal broadening,
which spreads the feature in velocity space and reduces its saturation. In
some cases, such as the $v\sim 725$\kms\ feature of sightline~\#1, the wings of
the dark matter feature become noticeably broader than those of the gas
features because of the higher temperatures at higher overdensities.

\subsection{The 1-D Flux Power Spectrum}\label{sec:fluxpk}
The \lyman\ flux power spectrum is a powerful tool for probing the dark
matter mass power spectrum on the smallest scales.  Because of the high
redshift, the continuous sampling of the line-of-sight density field,
and the moderate overdensity of absorbing structures, the \lya\ forest
provides a more direct link to the linear theory power spectrum than
other small-scale tracers.  (Primary cosmic microwave background
anisotropies are damped on these scales.)  However, to infer
information about the structure of the underlying dark matter, one must
understand the thermal structure of the IGM, as the amplitude and the
shape of the flux power spectrum are connected to the gas
temperature-density relation via equation~(\ref{eqn:tau}).  The bias of
the flux power spectrum, i.e., $b^2(k)=P_F(k)/P_{\rm lin}(k)$, is also
influenced by non-linear gravitational evolution, thermal broadening,
pressure support, peculiar velocities, and shock heating
\citep{croft98,croft99,croft02,viel04,viel08,mcdonald05}.

\begin{figure}
\includegraphics[angle=270,width=0.48\textwidth]{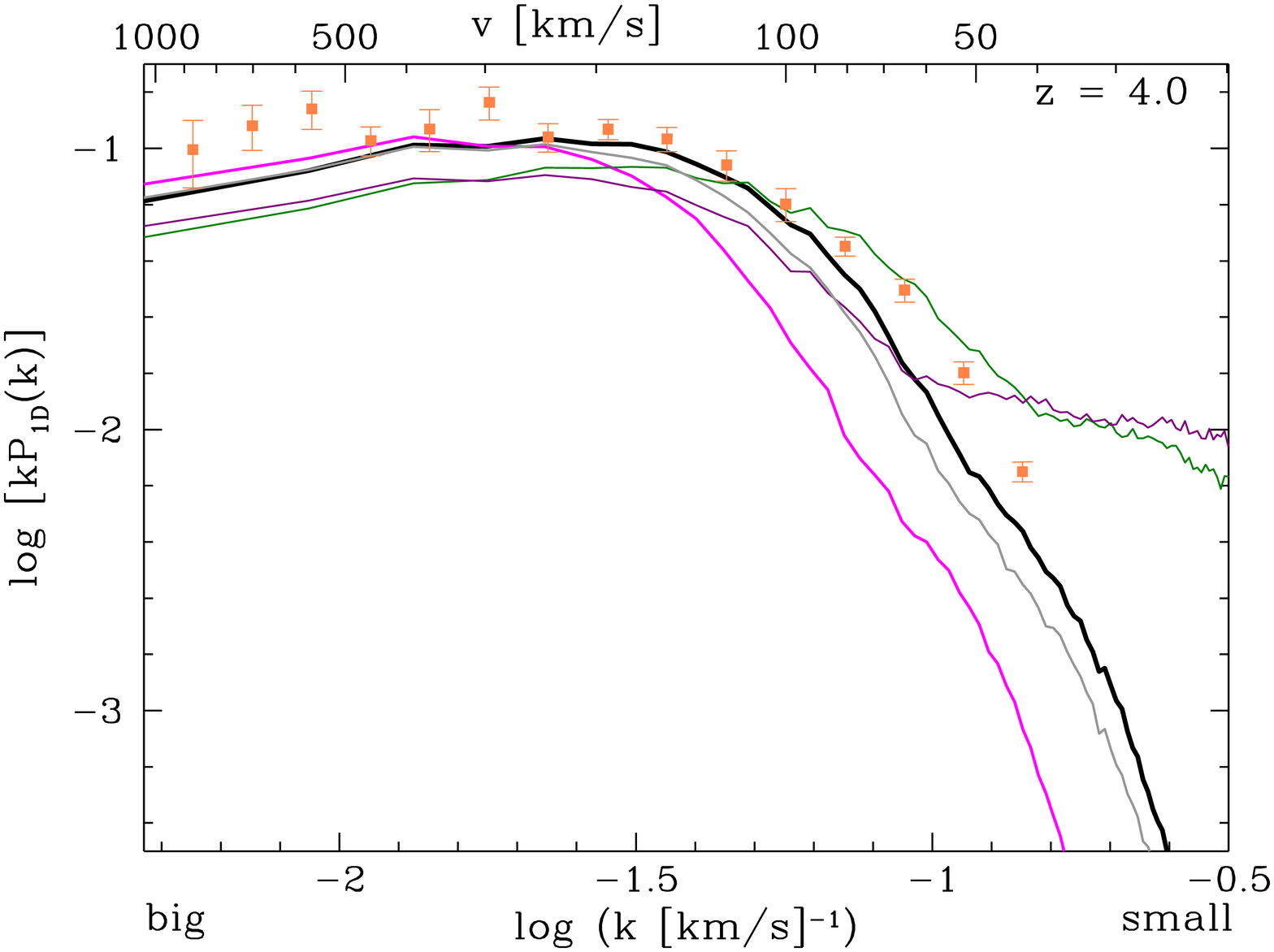}
\includegraphics[height=0.48\textwidth,angle=270]{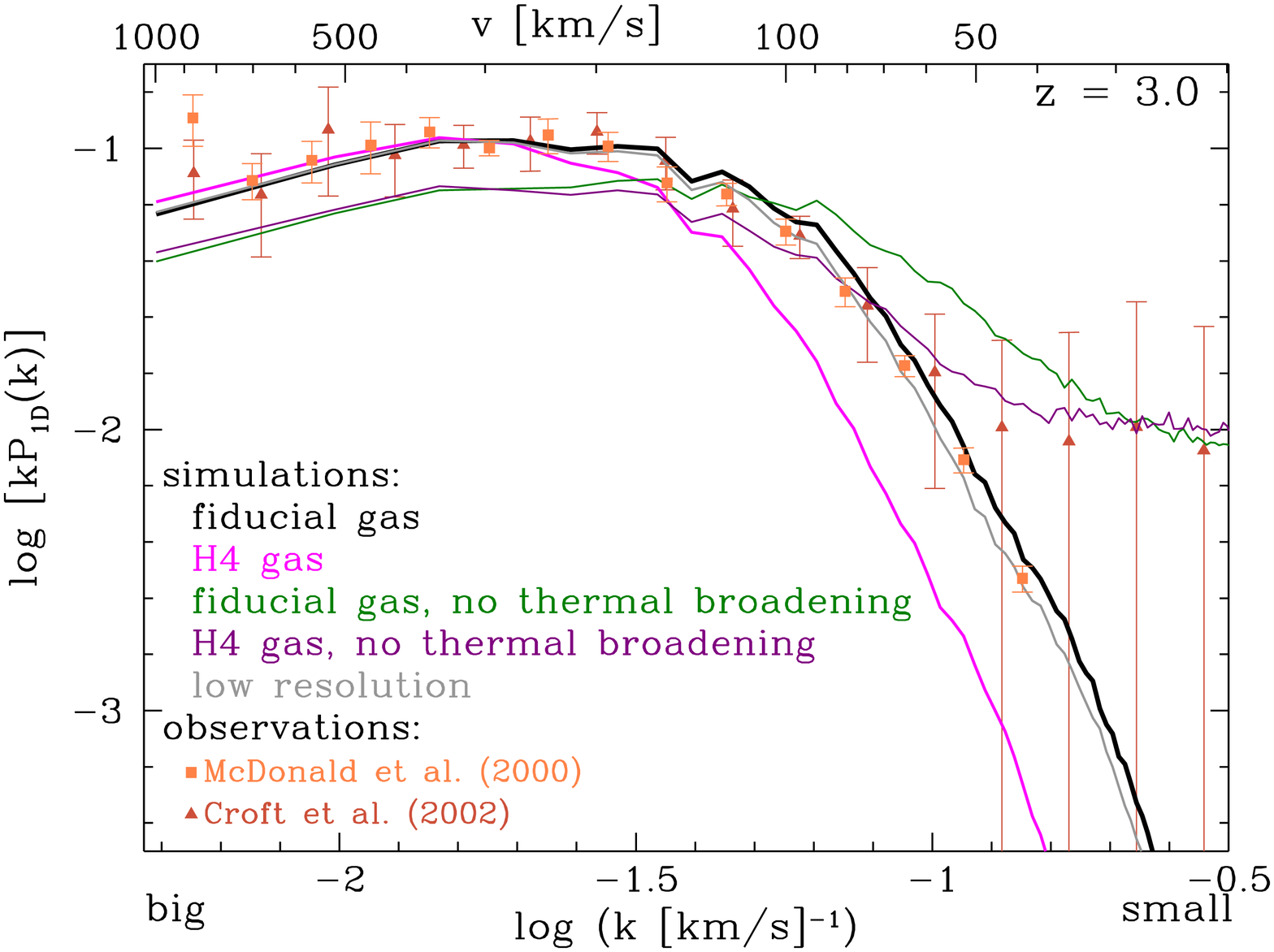}
\includegraphics[angle=270,width=0.48\textwidth]{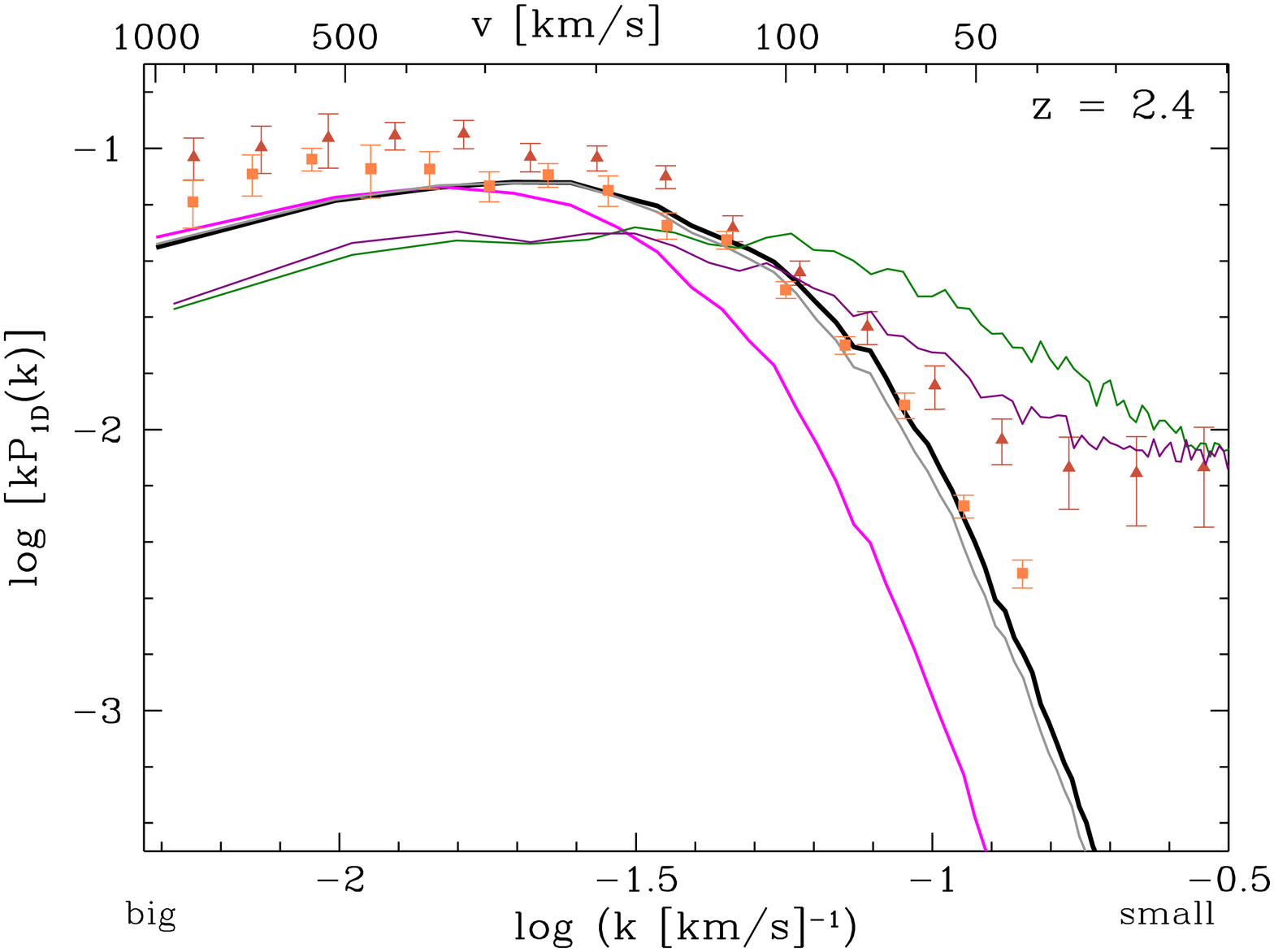}
\caption{\label{fig:fluxpk}The 1-D flux power spectrum at $z=4.0$ ({\em
  top}), $z=3.0$ ({\em middle}), and $z=2.4$ ({\em bottom}), with the
  fiducial simulation in black, the H4 simulation in pink, and the
  lower-resolution simulation in grey.  For reference, we also show the
  flux power spectrum in the absence of thermal broadening for the
  fiducial gas ({\em green}) and the H4 gas ({\em purple}), as well as
  the observed flux power spectrum from \citeauthor{mcdonald01}\
  (\citeyear{mcdonald01}, {\em squares}) and \citeauthor{croft02}\
  (\citeyear{croft02}, {\em triangles}).}
\end{figure}

Figure~\ref{fig:fluxpk} shows the one-dimensional line-of-sight flux
power spectrum, $P_{\mbox{\scriptsize 1D}}(k)$ of $(F-\langle
F\rangle)$, where $F\equiv\exp(-\tau_{\mbox{\tiny Ly}\alpha})$, based on
600 randomly selected sightlines through the fiducial, H4, and low
resolution simulations.  The flux power spectrum $P_{\mbox{\scriptsize
1D}}(k)$ is a measure of the variance of the flux, $\sigma_{\rm F}^2$,
on different scales; specifically, we adopt the normalization convention
of \cite{mcdonald00}, where
\begin{equation}\label{eqn:pk}
\sigma_F^2 = \int_{-\infty}^\infty \frac{{\rm d}k}{2\pi} P_{\mbox{\scriptsize 1D}}(k) =
\pi^{-1} \int_0^{\infty} {\rm d}k P_{\mbox{\scriptsize 1D}}(k).
\end{equation}
Also plotted in Figure~\ref{fig:fluxpk} are the flux power spectra for
the same lines of sight in both the fiducial and H4 simulations in the
absence of thermal broadening. For reference, we also plot the observed
flux power spectrum measured from high-resolution spectra given in
Table~4 of \citet[squares]{mcdonald00} and Table~7 (the ``B'' sample at
$z=2.4$ and the ``D'' sample at $z=3$) of \citet[triangles]{croft02}.
We have renormalized the \citet{croft02} measurements by
\citeauthor{mcdonald00}'s $\langle F\rangle^2$ to match our
normalization convention.

The cutoff in the flux power spectrum at large $k$ is a consequence of
thermal broadening, as is obvious from comparing the power spectra with
and without thermal broadening.  With thermal broadening included, the
low resolution ($144^3$ gas particles) and fiducial ($288^3$)
simulations produce very similar power spectra, indicating that even the
lower resolution simulation is well converged for this statistic.  At
high $k$, the power spectra of the H4 simulation are offset by roughly a
factor of $1.6$ in $k$ from the fiducial simulation, roughly consistent
with the difference of 2--$2.5$ in $T_0$ (since thermal velocities scale
as $T_0^{1/2}$).  The cutoff scale in the fiducial simulation agrees
better with the observational data, even though its temperatures are low
compared to the \citet{mcdonald01} estimates (see Fig.~\ref{fig:mcd01}).

Our cutoff scale also differs from that of \citet[][Figure~5]{lidz09};
the flux power spectrum of our fiducial simulation is more similar (on
small and large scales) to that of their hotter, $T_0=2\times 10^4$\,K
and $\alpha=0.3$, simulation than to their colder, $T_0=1\times 10^4$\,K
and $\alpha=0.6$, simulation, which has a \rhot\ relation similar to our
fiducial simulation.  We see no obvious reason for this discrepancy: the
resolution test in Figure~\ref{fig:fluxpk} shows good convergence, we do
not expect the cutoff scale to be sensitive to box size, we have checked
that high as density peaks produce the expected thermally broadened line
profiles in our extracted spectra, and we find good agreement between
the flux power spectrum from our $z=3$ spectra measured by our code
(written by R.\ Croft) and an independent code written by P.\
McDonald.\footnote{We thank Patrick McDonald for carrying out this test
for us.}  When spectra are extracted from our simulation using
\citeauthor{lidz09}'s code, the calculated $P(k)$ is consistent with the
one we measure;\footnote{We thank Adam Lidz for carrying out this test
for us.} the difference is therefore present in the simulated gas
distributions themselves.  For now, we draw no strong conclusions from
the comparison to data in Figure~\ref{fig:fluxpk}, and focus instead on
the relative roles of pressure support and thermal broadening.

\begin{figure}
\includegraphics[height=0.48\textwidth,angle=270]{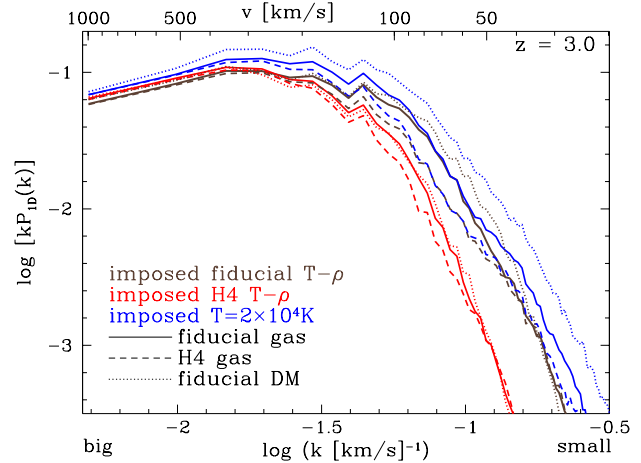}
\caption{\label{fig:fluxpkcomp}The 1-D flux power spectrum at $z=3$ for
  fiducial gas ({\em solid lines}), H4 gas ({\em dashed lines}), and
  fiducial dark matter ({\em dotted lines}) with fiducial ({\em
  brown}), H4 ({\em red}), and flat ({\em blue}) imposed
  temperature-density relations.}
\end{figure}

To separate the effects of gas pressure and thermal broadening,
Figure~\ref{fig:fluxpkcomp} shows $z=3$ flux power spectra from 200
randomly selected sightlines through the fiducial gas, H4 gas, and fiducial
dark matter density fields (indicated by the line type), each with three
imposed temperature-density relations:
fiducial, H4, and flat (indicated by the line color).
Comparing lines of the same color in
Figure~\ref{fig:fluxpkcomp} shows the effects of pressure support, with
the same \rhot\ relation applied to different density
distributions. Comparing lines of the same type but different colors
isolates the effect of the \rhot\ relation for the same underlying
density distribution.  The lines clearly separate into three groups
based on color, showing that thermal broadening dominates over pressure
support in determining the scale of the power spectrum cutoff.  There is
some difference among the three density distributions when we impose a
constant IGM temperature $T=2\times 10^4$\,K (blue lines), which shows
that some of the similarity for the other \rhot\ relations may reflect
the cancellation discussed in \S\,\ref{sec:compare}, where the
distribution with a larger Jeans length has less high density gas to
experience high thermal broadening.  However, on the whole our results
confirm the finding of \citet{mcdonald03} that it is the instantaneous
\rhot\ relation (at the epoch of observation) rather than the detailed
gas thermal history (and thus pressure history) that determines the
$P_F(k)$ cutoff.

\subsection{Flux Decrement Autocorrelation Functions}\label{sec:autocorr}
The flux decrement autocorrelation function, 
\begin{equation}\label{eqn:xiauto}
\xi_{\mbox{\scriptsize auto}}(\Delta v)\equiv\frac{\langle D(v)D(v+\Delta v)\rangle}{\langle D\rangle^2}, 
\end{equation}
is a commonly used tool for characterizing the \lya\ forest.  While
technically the one-dimensional flux power spectrum and flux
autocorrelation functions codify the same information ($P_F(k)$ is just
the one-dimensional Fourier transform of $\xi_{\mbox{\scriptsize
auto}}$), the flux autocorrelation function is often easier to describe
both observationally and from simulations because its definition does
not depend on information from all scales.  Furthermore, as
cross-correlation functions are more typically used to describe
information in closely paired lines of sight than cross-power spectra,
it is important for us to understand the effects of pressure and
temperature on the autocorrelation function before examining paired
sightlines (in Paper~II).

\begin{figure*}
\includegraphics[angle=270,width=\textwidth]{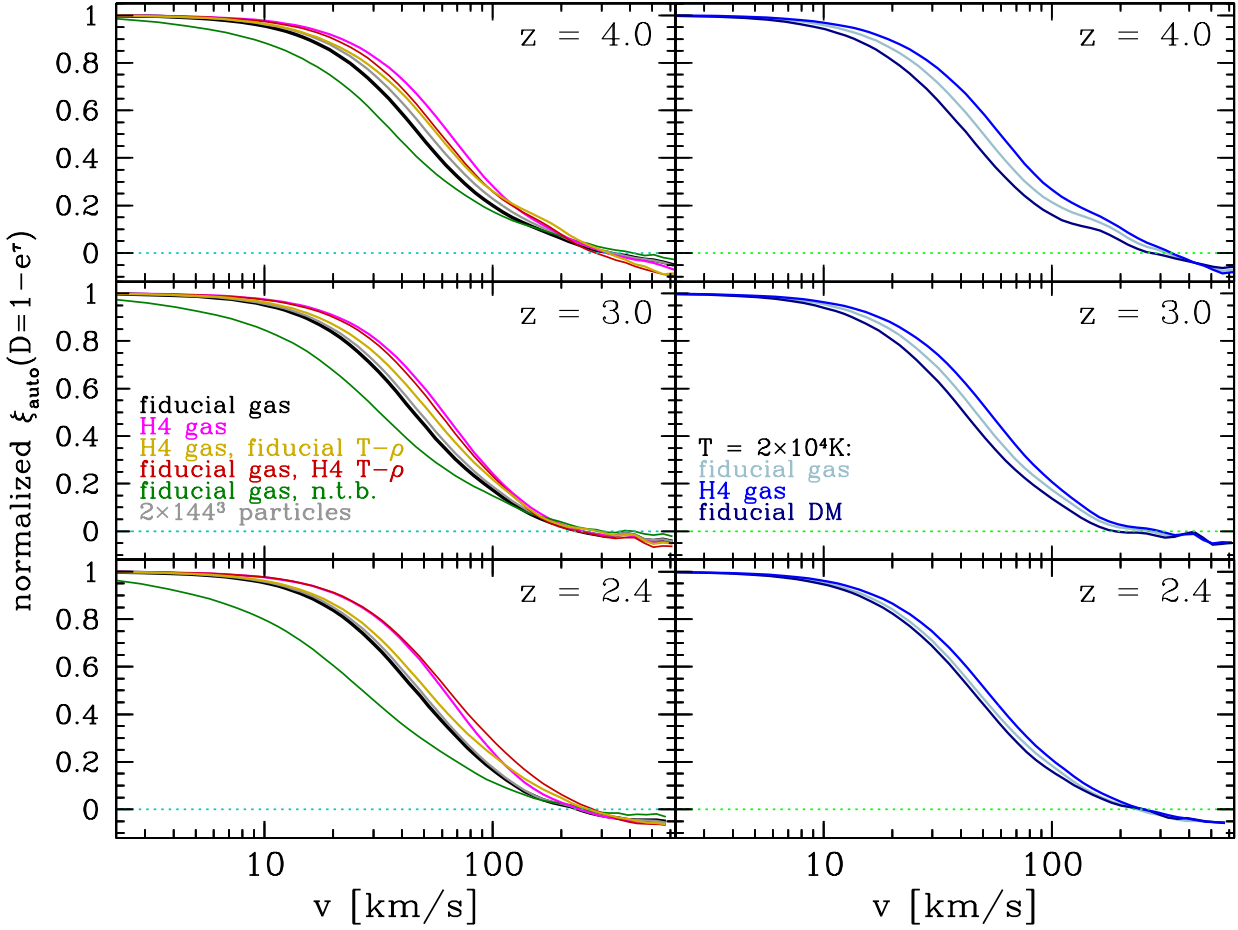}
\caption{\label{fig:autocorr}{\em Left}: Normalized flux decrement
autocorrelation functions, $\xi_{\mbox{\scriptsize auto}}\equiv\langle
D(v)D(v+\Delta v)\rangle/\langle D\rangle^2$, at $z=2.4,\,3.0$, and 4.0.
The black curve with error bars shows results for the fiducial
simulation, while the pink and grey lines show results for the H4 and
low-resolution simulations, respectively.  Red and mustard lines show,
respectively, the fiducial gas with the H4 \rhot\ relation imposed and
the H4 gas with the fiducial \rhot.  The green line shows the fiducial
gas with no thermal broadening. {\em Right}: Normalized
$\xi_{\mbox{\scriptsize auto}}$, for the fiducial gas ({\em light
blue}), H4 gas ({\em bright blue}), and fiducial dark matter ({\em dark
blue}), all with an imposed temperature of $T=2\times10^4$\,K.  }
\end{figure*}

Figure~\ref{fig:autocorr} plots normalized flux decrement
autocorrelation functions at $z=2.4$, 3.0, and 4.0.  We normalize
$\xi_{\rm auto}=1$ at the smallest scales to highlight differences in
the turnover scale of the correlation function rather than differences
in normalization.  The fiducial and low-resolution simulations have
nearly identical $\xi_{\rm auto}$, while the H4 simulation has a
noticeably larger coherence scale.  However, if we impose the fiducial
\rhot\ relation on the H4 gas, then $\xi_{\rm auto}$ is nearly identical
to that of the fiducial simulation.  Conversely, imposing the H4 \rhot\
on the fiducial simulation yields nearly the same $\xi_{\rm auto}$ as
the H4 simulation.\footnote{If we impose the fiducial (H4) \rhot\
relation on the fiducial (H4) simulation, instead of using the
simulation temperatures themselves, then the change in $\xi_{\rm auto}$
is negligible.}

While these results suggest that the effects of pressure are small
compared to those of thermal broadening, the right-hand panels of
Figure~\ref{fig:autocorr} show that $\xi_{\rm auto}$ is different for
the H4 gas, fiducial gas, and fiducial dark matter if we impose a flat
$T=2\times 10^4$\,K on all particles.  In this case, the higher pressure
distribution exhibits a larger coherence scale, especially at high
redshifts.  Thus, the similarity of the fiducial and H4 distributions
with the same imposed \rhot, seen in the left-hand panels, arises partly
from the counter-balancing effects of $T_0$ and the changes in the
distribution of gas densities discussed in \S\,\ref{sec:compare} (see
Figure~\ref{fig:rhohist}).

\subsection{Flux Decrement Probability Distributions}\label{sec:pd}
The flux decrement probability distribution function (PDF) is a
potentially powerful tool for probing the IGM, both by itself and as a
means of adding constraints to other statistical measures such as the
flux power spectrum
\citep{rauch97,weinberg99,gaztanaga99,nusser00,desjacques05,desjacques07}.
In particular, \citet{bolton08} have recently used the \lya\ forest flux
distribution as measured by \citet{kim07} to suggest that the
temperature-density relation at $z\sim 3$ is {\em inverted} ($\alpha <
0$ in equation~\ref{eqn:rhot}), with low-density gas at higher
temperature than higher density gas.  While the differences owing to
changes in the \rhot\ relation are not as extreme as those in $P_F(k)$
or $\xi_{\mbox{\scriptsize auto}}$, the flux decrement PDF can
potentially be measured to much higher accuracy using the same number of
sightlines.  On the other hand, the PDF is sensitive to continuum
fitting, while on small scales $P_F(k)$ is not (on large scales
continuum fitting can be problematic even for $P_F(k)$; e.g.,
\citealt{kim04}).  Furthermore, high-resolution spectra are needed to
accurately measure the PDF \citep{viel04,tytler04}. Since
high-resolution spectra will always exist in smaller numbers than
low-resolution spectra, it should be noted that the PDF is also
potentially sensitive to sample variance, the explanation \citet{kim07}
suggest for the $\sim$30\% difference between their $z\sim3$ PDF and the
one measured by \citet{mcdonald00} on a very similar data set.

\begin{figure}
\includegraphics[width=0.48\textwidth]{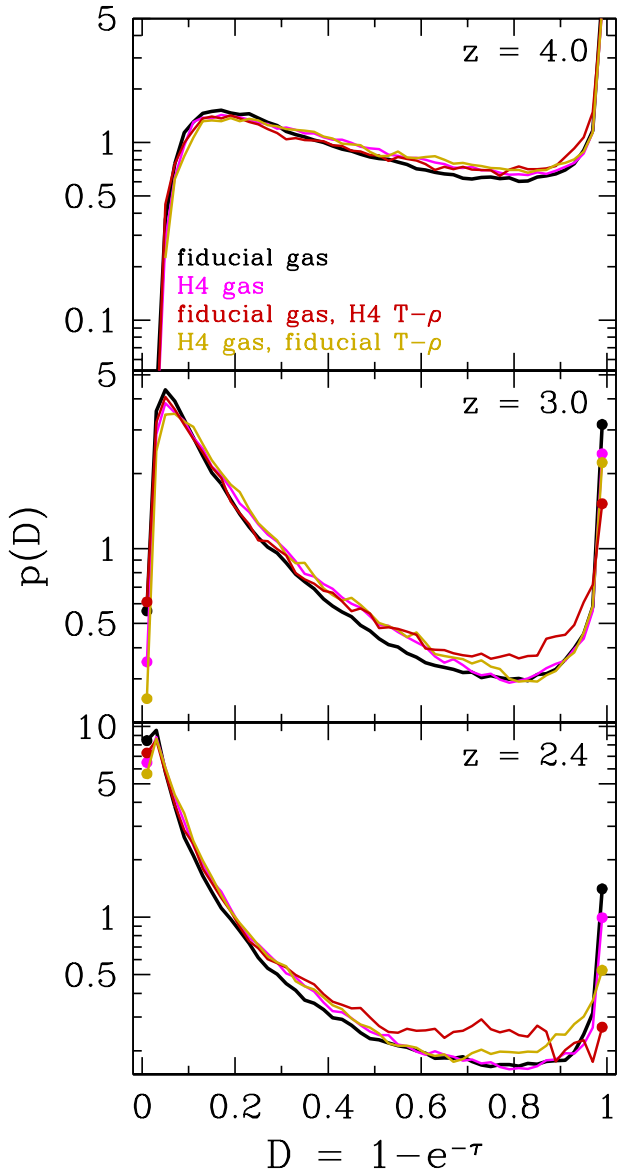}
\caption{\label{fig:PDzall}Flux decrement probability distributions in
  linear bins of $D$ at $z=2.4$, 3, and 4 for the fiducial and H4
  simulations, with ({\em black} and {\em pink}) inherent temperatures
  and imposed H4 ({\em red}) and fiducial ({\em mustard}) \rhot\
  relations, respectively. The points at $z=2.4$ and 3 are included to
  show the $p(D)$ in the most transparent ({\em left}) and
  opaque ({\em right}) bins.}
\end{figure}

\begin{figure}
\includegraphics[width=0.48\textwidth]{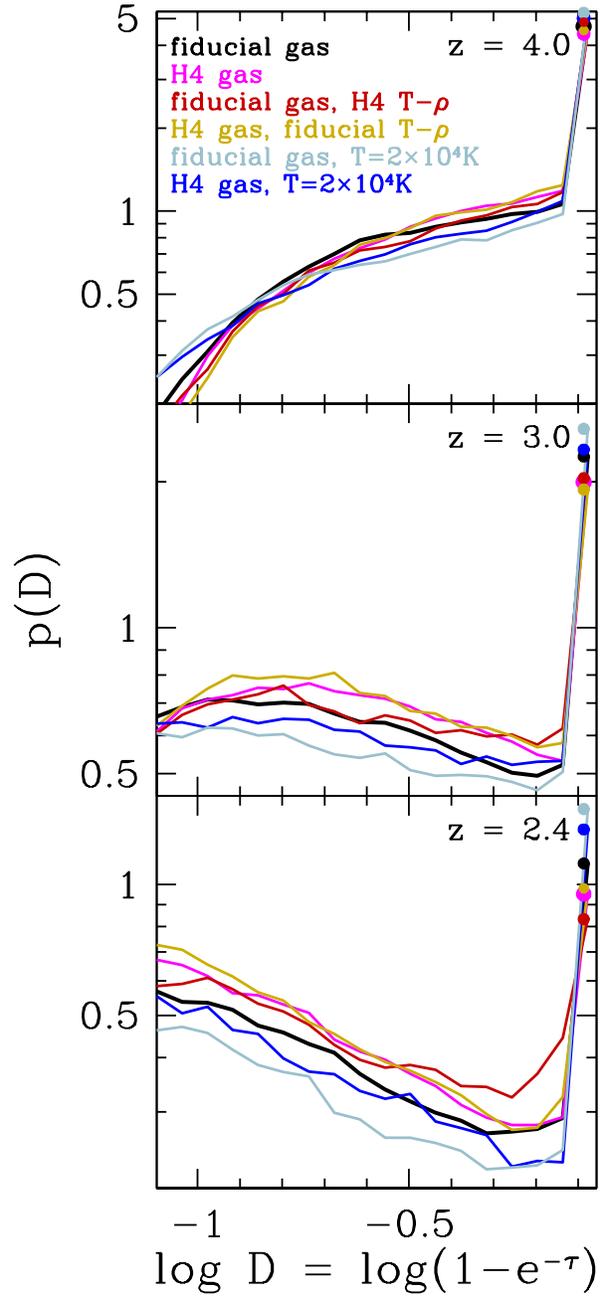}
\caption{\label{fig:PDlogzall}Flux decrement probability distributions
  in logarithmic bins of $D$ at $z=2.4$, 3, and 4 for the fiducial gas
  distribution, with inherent temperatures ({\em black}), at low
  resolution ({\em grey}), and with imposed H4 \rhot\ ({\em red}).  Also
  shown are the imposed H4 gas distribution with inherent temperatures
  ({\em pink}), with the imposed fiducial \rhot\ ({\em mustard}), as
  well as both the fiducial and H4 gas distributions with an imposed
  flat $T=2\times 10^4$\,K ({\em light blue} and {\em bright blue},
  respectively).  The points are included to show the $p(D)$ in the
  final (most opaque) bin.}
\end{figure}

Figure~\ref{fig:PDzall} shows the linear flux decrement PDF, i.e.,
$p(D)$, where $p(D)\Delta D$ is the number of pixels in a bin of width
$\Delta D$ divided by the total number of pixels.  In
Figure~\ref{fig:PDlogzall} we show for a larger set of models the
logarithmic PDF, $p(\log D)$, which allows better visual discrimination
in the range $0.1 < D < 0.5$.  The shape of the PDF changes radically
with redshift as the mean flux decrement changes; physically, this
change is driven by the mean density dropping as $(1+z)^3$.  However, as
all the models at a given redshift are normalized to the same $\langle
D\rangle$, the effects of \rhot\ changes on the {\em mean} absorption
are removed.  Though for visual clarity we do not plot the PDFs from the
$2\times 144^3$ particle simulation, the low resolution and fiducial
PDFs differ by $\sim$1--5\%; this difference is much smaller than the
typical model differences in Figures~\ref{fig:PDzall} and
\ref{fig:PDlogzall}.

Thermal broadening reduces the number of transparent ($D\approx 0$)
pixels, while pressure support increases the number of pixels at extreme
flux decrements (for visual clarity, we do not plot the dark matter and
non-thermally broadened PDFs).  At all redshifts, the spectra
generated from the full H4 simulation have more pixels with mid-range
flux decrements, i.e. relative to the fiducial simulation, the H4
simulation produces fewer opaque pixels and fewer transparent pixels.
This is as expected, since higher pressure smooths the gas distribution
and higher temperature leads to greater thermal broadening.  However,
the fiducial simulation with the imposed H4 \rhot\ has fewer saturated
pixels than the H4 simulation itself, and more pixels with $D\approx
0.6$--$0.8$.  This surprising result highlights the sometimes
complicated interplay between the density distribution and the \rhot\
relation.  The fiducial simulation has more high overdensity gas
(Figure~\ref{fig:rhohist}), but because this gas is at high temperature
(with $T\propto[1+\delta]^{0.6}$), thermal broadening converts narrow,
fully saturated features to broader, moderately saturated ones.

The slope of the \rhot\ relation has a direct impact on the flux PDF,
independent of thermal broadening, because it affects the mapping from
density contrast to flux.  Equation~(\ref{eqn:tau}) gives
$\tau_{\mbox{\tiny HI}}\propto(1+\delta)^{2-0.7\alpha}$, so
$\tau_{\mbox{\tiny HI}}\propto(1+\delta)^{1.6}$ for our fiducial and H4
relations (with $\alpha\approx 0.6$) and $\tau_{\mbox{\tiny
HI}}\propto(1+\delta)^{2}$ for our flat, $T=2\times 10^4$\,K relation
($\alpha=0$).  Imposing a constant $T$ substantially alters the PDFs of
both the fiducial and H4 simulations (blue curves in
Figure~\ref{fig:PDlogzall}), with the greater sensitivity of
$\tau_{\mbox{\tiny HI}}$ to $(1+\delta)$ leading to more saturated
pixels and fewer mid-range pixels.  The two simulations have
significantly different PDFs, which with constant thermal broadening
must arise from effects of pressure on the gas density distribution.
In contrast to the power spectrum and flux correlation function, thermal
broadening and pressure support have comparable impact on the flux PDF,
and they interact in complex ways.  This different behavior arises
because the PDF responds directly to the full overdensity distribution
and its mapping to flux, while the power spectrum and correlation
function measure the variance of this distribution as a function of
scale.

\section{Conclusions}\label{sec:conc}
We have investigated the relative importance of pressure support and
thermal broadening in determining the longitudinal structure of the
\lyman\ forest.  Our main results come from comparing two SPH
simulations with identical initial conditions but different
photoionization heating rates, which produce a factor of $\sim 2.3$
difference in the temperature of diffuse IGM gas.  We have imposed
different temperature-density relations on the simulation outputs to
isolate physical effects, extracted spectra from the dark matter
distribution to extend our investigation to the pressureless case, and
compared the fiducial simulation ($2\times 288^3$ particles) to a lower
resolution simulation ($2\times 144^3$) to quantify numerical resolution
effects.

Equation~(\ref{eqn:vjsigth}) shows that the Hubble flow across the Jeans
scale is generally of the same magnitude as thermal broadening
($H\lambda_J\sim\sigma_{\rm th}$).  However, the IGM is an expanding,
inhomogeneous medium evolving in the potential of a non-linear dark
matter distribution, so the Jeans length is at best an approximate
description of the scale imposed by gas pressure support.  It is
therefore difficult to know without detailed simulations whether thermal
broadening or pressure support will dominate the structure of the
forest.  A related question is the meaning of the density contrast
$\delta$ in the fluctuating Gunn-Peterson approximation
(equation~\ref{eqn:tau}).  In principle, this should be the density of
the gas that is absorbing \lya\ photons, but analyses of early SPH
simulations found that \lya\ forest spectra created from the
(pressureless) dark matter distribution were remarkably similar to those
created from the (pressure supported) gas distribution
\citep[e.g.,][]{croft98}. However, the relatively low resolution of
those simulations (a mass resolution that is $\sim 60$ times lower than
our fiducial simulation here) raised the possibility that both sets of
spectra were artificially broadened to the numerical resolution limit.
We investigate this issue more confidently here because the initial
particle spacing of our $288^3$ simulations, $43\,h^{-1}$\,kpc comoving,
is far below the typical Jeans length at IGM overdensities,
$\lambda_J\sim\, 800h^{-1}$\,kpc comoving, and because our $144^3$
simulation allows a direct resolution test.

In broad brush, our conclusions are that thermal broadening dominates
over pressure support in determining the visual appearance and
statistical properties of the \lya\ forest, but that differences in gas
pressure do have a noticeable effect.  The widths of absorption features
are typically  set by Hubble flow across the absorbing structure, though
thermal broadening does smooth out small scale corrugations to create
coherent features (see Fig.~\ref{fig:zoo}).  Once we include thermal
broadening, the spectra created from dark matter distributions are similar
to those created from the gas, even though the effects of pressure
support on the gas density field are readily discernible.  Thus, one can
drastically change the Jeans scale without drastically changing the
forest.  However, dark matter spectra are visually and statistically
distinguishable from gas spectra, more so than in the earlier generation
of lower resolution SPH simulations.

Turning to individual statistics, we find that thermal broadening sets
the turnover scale of the one-dimensional flux power spectrum, with the
fiducial gas, H4 gas, and dark matter distributions producing similar
power spectra if one imposes the fiducial or H4 \rhot\ relation on all
three.  Similar conclusions hold for the coherence scale of the flux
decrement autocorrelation function.  However, in both cases, the weak
impact of pressure support partly owes to a cancellation effect that
arises with a sloped \rhot\ relation: a higher pressure distribution has
more ``Jeans broadening,'' but it has less thermal broadening because
there is less high overdensity, high temperature gas.  When we impose a
constant IGM temperature of $T=2\times 10^4$\,K, the differences among
the three cases are more noticeable, though they are still small
compared to the effects of thermal broadening.  For the flux decrement
probability distribution function, thermal broadening and pressure
support have effects of comparable magnitude, though thermal broadening
is still somewhat more important.

Our $144^3$ and $288^3$ simulations yield similar results for all our
statistics.  The resolution effects are larger at $z=4$ than at lower
redshifts, where they have a small but noticeable impact on all the
statistics.  For most purposes, the resolution of our $144^3$
simulations (in a $12.5\,h^{-1}$\,Mpc~comoving volume) is adequate.

Though it is a stronger player than pressure support in setting the
scale of the longitudinal \lya\ forest, thermal broadening is an
inherently one-dimensional phenomenon.  Because pressure acts in three
dimensions, we expect it to play the main role in setting the {\em
transverse} coherence of the \lya\ forest across neighboring sightlines.
Growing samples of binary quasars with separations of $\Delta\theta\lesssim
10$\arcsec\ now make it possible to probe the expected Jeans scale
\citep{hennawi06,hennawi09}.  We show in Paper~II that the degree of
transverse coherence on these scales is indeed sensitive to gas pressure
support and insensitive to thermal broadening.  Observational studies of
close quasar pairs can directly probe the scale of pressure support on
the \lyman\ forest, providing new insights into the physical state and
thermal history of the high-redshift intergalactic medium.

\section*{Acknowledgments}

We gratefully acknowledge Joe Hennawi and Eduardo Rozo for helpful
discussions and comments on earlier drafts.  We thank Pat McDonald,
Rupert Croft, Adam Lidz, and Matias Zaldarriaga for their help in
sorting through the power spectrum normalization conventions in
different observational and theoretical analyses.  We are grateful to
the anonymous referee for thoughtful suggestions on the text.  This work
has been supported in part by NSF grant AST-0707985 and NASA ADP grant
NNX08AJ44G.

\label{lastpage}

\end{document}